%
\input phyzzx
\input phyzzx.plus
\overfullrule=0pt
\def\tr{\mathop{\rm tr}\nolimits}
\def\str{\mathop{\rm str}\nolimits}
\def\Tr{\mathop{\rm Tr}\nolimits}
\def\Det{\mathop{\rm Det}\nolimits}
\def\Ln{\mathop{\rm Ln}\nolimits}
\def\sym{\mathop{\rm sym}\nolimits}
\def\sqr#1#2{{\vcenter{\hrule height.#2pt
      \hbox{\vrule width.#2pt height#1pt \kern#1pt
          \vrule width.#2pt}
      \hrule height.#2pt}}}
\def\square{{\mathchoice{\sqr84}{\sqr84}{\sqr{5.0}3}{\sqr{3.5}3}}}
\def\dA{\mathop\square\nolimits}
\def\QED{\hfil$\dA$\break}
\REF\NIE{H. B. Nielsen and M. Ninomiya,
Phys.\ Lett.\ {\bf B105} (1981) 219; Nucl.\ Phys.\ {\bf B185} (1981)
20; {\bf B195} (1982) 541~(E); {\bf B193} (1981) 173.}
\REF\FRI{D. Friedan,
Commun.\ Math.\ Phys.\ {\bf 85} (1982) 481.}
\REF\ADL{S. L. Adler,
Phys.\ Rev.\ {\bf 177} (1969) 2426.}
\REF\BEL{J. S. Bell and R. Jackiw,
Nuovo Cim.\ {\bf 60A} (1969) 47.}
\REF\BAR{W. A. Bardeen,
Phys.\ Rev.\ {\bf 184} (1969) 1848.}
\REF\BOU{C. Bouchiat, J. Iliopoulos and Ph.\ Meyer,
Phys.\ Lett.\ {\bf 38B} (1972) 519.}
\REF\GEO{H. Georgi and S. Glashow,
Phys.\ Rev.\ {\bf D6} (1972) 429.}
\REF\GRO{D. Gross and R. Jackiw,
Phys.\ Rev.\ {\bf D6} (1972) 477.}
\REF\SHA{For a review on various approaches, Y. Shamir,
Nucl.\ Phys.\ Proc.\ Suppl.\ {\bf 47} (1996) 212.}
\REF\BEC{C. Becchi, A. Rouet and R. Stora,
Comm.\ Math.\ Phys.\ {\bf 42} (1975) 127; Ann.\ Phys.\ {\bf 98}
(1976) 287.}
\REF\STO{R. Stora,
in {\sl New developments in quantum field theory and statistical
mechanics\/} (Carg\`ese 1976), eds.\ M. L\'evy and P. Mitter,
(Plenum Press, New York, 1977).}
\REF\BON{L. Bonora and P. Cotta-Ramusino,
Phys.\ Lett.\ {\bf 107B} (1981) 87; Commun.\ Math.\ Phys.\ {\bf 87}
(1983) 589.}
\REF\STOR{R. Stora,
in {\sl Progress in gauge field theory\/} (Carg\`ese 1983),
eds.\ G. 't Hooft et al, (Plenum Press, New York, 1984).}
\REF\ZUM{B. Zumino,
in {\sl Relativity, groups and topology\/} (Les Houches 1983),
eds.\ B. S. De Witt and R. Stora, (North-Holland, Amsterdam,
1984).\nextline
B. Zumino, Y. S. Wu and A. Zee,
Nucl.\ Phys.\ {\bf B239} (1984) 477.}
\REF\BAU{L. Baulieu,
in {\sl Particles and fields\/} (Carg\`ese 1983), eds.\ M. L\'evy
et al, (Plenum Press, New York, 1985);
Nucl.\ Phys.\ {\bf B241} (1984) 557; Phys.\ Rep.\ {\bf 129} (1985) 1.}
\REF\BRA{F. Brandt, N. Dragon and M. Kreuzer,
Phys.\ Lett.\ {\bf B231} (1989) 263; Nucl.\ Phys.\ {\bf B332} (1990)
224; {\bf B332} (1990) 250.\nextline
N. Dragon,
Lectures given at Saalburg Summer School (1995), hep-th/9602163.}
\REF\DUB{M. Dubois-Violette, M. Henneaux, M. Talon and C.-M. Viallet,
Phys.\ Lett.\ {\bf B267} (1991) 81; {\bf B289} (1992) 361.}
\REF\BOR{A. Borrelli, L. Maiani, G. C. Rossi, R. Sisto and M. Testa,
Phys.\ Lett.\ {\bf B221} (1989) 360; Nucl.\ Phys.\ {\bf B333} (1990)
335.}
\REF\LUS{M. L\"uscher,
Nucl.\ Phys.\ {\bf B538} (1999) 515.}
\REF\LUSC{M. L\"uscher,
Nucl.\ Phys.\ {\bf B549} (1999) 295.}
\REF\LUSCH{M. L\"uscher,
Nucl.\ Phys.\ {\bf B568} (2000) 162; for a review,
Nucl.\ Phys.\ Proc.\ Suppl.\ {\bf 83-84} (2000) 34.}
\REF\WES{J. Wess and B. Zumino,
Phys.\ Lett.\ {\bf 37B} (1971) 95.}
\REF\GIN{P. H. Ginsparg and K. G. Wilson,
Phys.\ Rev.\ {\bf D25} (1982) 2649.}
\REF\HAS{P. Hasenfratz,
Nucl.\ Phys.\ Proc.\ Suppl.\ {\bf 63A-C} (1998) 53;
Nucl.\ Phys.\ {\bf B525} (1998) 401.}
\REF\NEU{H. Neuberger,
Phys.\ Lett.\ {\bf B417} (1998) 141; {\bf B427} (1998) 353.}
\REF\NAR{R. Narayanan and H. Neuberger,
Phys.\ Rev.\ Lett. {\bf 71} (1993) 3251; Nucl.\ Phys.\ {\bf B412}
(1994) 574; {\bf B443} (1995) 305.}
\REF\RAN{S. Randjbar-Daemi and J. Strathdee,
Phys.\ Lett.\ {\bf B348} (1995) 543; Nucl.\ Phys.\ {\bf B443} (1995)
386; {\bf B466} (1996) 335; Phys.\ Lett.\ {\bf B402} (1997) 134.}
\REF\CON{A. Connes,
{\sl Noncommutative geometry\/} (Academic Press, New York, 1994).}
\REF\SIT{A. Sitarz,
J. Geom.\ Phys.\ {\bf 15} (1995) 123.}
\REF\DIM{A. Dimakis and F. M\"uller-Hoissen,
Phys.\ Lett.\ {\bf B295} (1992) 242;
J. Phys. A: Math.\ Gen.\ {\bf 27} (1994) 3159;
J.\ Math.\ Phys.\ {\bf 35} (1994) 6703.\nextline
A. Dimakis, F. M\"uller-Hoissen and T. Striker, 
Phys.\ Lett.\ {\bf B300} (1993) 141;
J. Phys. A: Math.\ Gen.\ {\bf 26} (1993) 1927.}
\REF\DIN{H. G. Ding, H. Y. Guo, J. M. Li and K. Wu,
Z. Phys.\ {\bf C64} (1994) 521; J. Phys.\ A: Math.\ Gen.\ {\bf 27}
(1994) L75; {\bf 27} (1994) L231; Commun.\ Theor.\ Phys.\ {\bf 21}
(1994) 85.\nextline
H. Y. Guo, K. Wu and W. Zhang,
``Noncommutative differential calculus on discrete abelian groups and
its applications,'' ITP-Preprint, March, 1999.}
\REF\FUJ{T. Fujiwara, H. Suzuki and K. Wu,
Nucl.\ Phys.\ {\bf B569} (2000) 643.}
\REF\FUJI{T. Fujiwara, H. Suzuki and K. Wu,
Phys.\ Lett.\ {\bf B463} (1999) 63; hep-lat/9910030.}
\REF\GOC{M. G\"ockeler, A. S. Kronfeld, M. L. Laursen, G. Schierholz
and U.-J. Wiese,
Nucl.\ Phys.\ {\bf B292} (1987) 349.\nextline
M. G\"ockeler, A. S. Kronfeld, G. Schierholz and U.-J. Wiese,
Nucl.\ Phys.\ {\bf B404} (1993) 839.}
\REF\HER{P. Hern\'andez and R. Sundrum,
Nucl.\ Phys.\ {\bf B455} (1995) 287; {\bf B472} (1996) 334.}
\REF\REE{M. Reed and B. Simon,
{\sl Functional analysis} (Academic Press, New
York, 1972).}
\REF\HERN{P. Hern\'andez, K. Jansen and M. L\"uscher,
Nucl.\ Phys.\ {\bf B552} (1999) 363.}
\REF\HASE{P. Hasenfratz, V. Laliena and F. Niedermayer,
Phys.\ Lett.\ {\bf B427} (1998) 125.}
\REF\LUSCHE{M. L\"uscher,
Phys.\ Lett.\ {\bf B428} (1998) 342.}
\REF\LUSCHER{M. L\"uscher,
Commun.\ Math.\ Phys.\ {\bf 85} (1982) 39.}
\REF\NIED{F. Niedermayer,
Nucl.\ Phys.\ Proc.\ Suppl.\ {\bf 73} (1999) 105.}
\REF\HOR{I. Horvath,
Phys.\ Rev.\ Lett.\ {\bf 81} (1998) 4063; Phys.\ Rev.\ {\bf D60}
(1999) 034510.}
\REF\BIE{W. Bietenholz,
hep-lat/9901005.}
\REF\FUJIW{T. Fujiwara, H. Suzuki and K. Wu,
hep-lat/0001029.}
\REF\AOY{T. Aoyama and Y. Kikukawa,
hep-lat/9905003.}
\REF\NEUB{H. Neuberger,
hep-lat/9912013.}
\REF\NARA{R. Narayanan,
Phys.\ Rev.\ {\bf D58} (1998) 097501.}
\REF\KIK{Y. Kikukawa and A. Yamada,
Nucl.\ Phys.\ {\bf B547} (1999) 413.}
\REF\SUZ{H. Suzuki,
Prog.\ Theor.\ Phys.\ {\bf 101} (1999) 1147.}
\REF\BAER{O. B\"ar and I. Campos,
Nucl.\ Phys.\ Proc.\ Suppl.\ {\bf 83-84} (2000) 594; hep-lat/0001025,
to appear in Nucl.\ Phys.\ B.}
\REF\WIT{E. Witten,
Phys.\ Lett.\ {\bf 117B} (1982) 324.}
\REF\FUJIK{K. Fujikawa,
Nucl.\ Phys.\ {\bf B546} (1999) 480.}
\REF\KIKU{Y. Kikukawa and A. Yamada,
Phys.\ Lett.\ {\bf B448} (1999) 265.}
\REF\ADA{D. H. Adams,
hep-lat/9812003.}
\REF\SUZU{H. Suzuki,
Prog.\ Theor.\ Phys.\ {\bf 102} (1999) 141.}
\REF\BARD{W. Bardeen and B. Zumino,
Nucl.\ Phys.\ {\bf B244} (1984) 421.}
\REF\NEUBE{H. Neuberger,
Phys.\ Lett. {\bf B437} (1998) 117; Phys.\ Rev.\ {\bf D59} (1999)
085006.}
\REF\FER{S. Ferrara, O. Piguet and S. Schweda,
Nucl.\ Phys.\ {\bf B119} (1977) 493.}
\REF\FUJIKA{K. Fujikawa,
Prog.\ Theor.\ Phys.\ {\bf 59} (1978) 2045.}
\REF\BONO{L. Bonora and M. Tonin,
Phys.\ Lett.\ {\bf 98B} (1981) 48.}
\REF\FAD{L. D. Faddeev,
Phys.\ Lett.\ {\bf 145B} (1984) 81.}
\REF\ORA{See, for example, L. O'Raifeartaigh,
{\sl Group structure of gauge theories\/} (Cambridge University Press,
Cambridge, 1986).}
\REF\AOK{S. Aoki,
Phys.\ Rev.\ {\bf D35} (1986) 1435.}
\REF\COS{A. Coste, C. Korthals Altes and O. Napoly,
Phys.\ Lett.\ {\bf 179B} (1986) 125; Nucl.\ Phys.\ {\bf B289} (1987)
645.}
\REF\ALV{L. Alvarez-Gaume and P. Ginsparg,
Nucl.\ Phys.\ {\bf B243} (1984) 449.}
\REF\ADAM{D. H. Adams,
hep-lat/9910036; hep-lat/0001014.}
\REF\FOR{D. Foerster, H. B. Nielsen and M. Ninomiya,
Phys.\ Lett.\ {\bf 94B} (1980) 135.}
\REF\GOL{M. Golterman and Y. Shamir,
Phys.\ Lett.\ {\bf B353} (1995) 84; {\bf B359} (1995) 422~(E);
Nucl.\ Phys.\ Proc.\ Suppl.\ {\bf 47} (1996) 603.}
\Pubnum={IC/2000/8\cr IU-MSTP/40\cr hep-lat/0002009}
\date={\null}
\titlepage
\title{Anomaly Cancellation Condition in Lattice Gauge Theory}
\author{Hiroshi Suzuki\foot{E-mail: hsuzuki@ictp.trieste.it}}
\address{Department of Mathematical Sciences,
Ibaraki University\break
Mito 310-8512, Japan\foot{\rm On leave of absence from.}}
\andaddress{The Abdus Salam International Center for Theoretical
Physics, Trieste, Italy}
\abstract{%
We study the gauge anomaly~${\cal A}$ defined on a 4-dimensional
infinite lattice while keeping the lattice spacing finite. We assume
that (I)~${\cal A}$~depends smoothly and locally on the gauge
potential, (II)~${\cal A}$~reproduces the gauge anomaly in the
continuum theory in the classical continuum limit, and
(III)~$U(1)$~gauge anomalies have a topological property. It is then
shown that the gauge anomaly~${\cal A}$ can always be removed by local
counterterms to all orders in powers of the gauge potential, leaving
possible breakings proportional to the anomaly in the continuum
theory. This follows from an analysis of nontrivial local solutions to
the Wess-Zumino consistency condition in lattice gauge theory. Our
result is applicable to the lattice chiral gauge theory based on the
Ginsparg-Wilson Dirac operator, when the gauge field is sufficiently
weak~$\|U(n,\mu)-1\|<\epsilon'$, where $U(n,\mu)$~is the link variable
and $\epsilon'$~a certain small positive constant.}
\bigskip
\noindent
PACS numbers: 11.15.Ha, 11.30.Rd\hfill\break
\endpage
\chapter{Introduction}
If one puts Weyl fermions on a lattice while respecting desired
physical properties, one has to sacrifice the
$\gamma_5$-symmetry~[\NIE,\FRI]. This implies that the gauge symmetry
is inevitably broken on the lattice when Weyl fermions are coupled to
the gauge field. This is rather natural, because we know that the
gauge anomaly exists in the continuum theory~[\ADL--\GRO]. However,
even if the anomaly in the continuum theory cancels,
$\tr_{R-L}T^a\{T^b,T^c\}=0$~[\BAR--\GRO], the fermion determinant is
not gauge invariant in general when the lattice spacing is finite,
$a\neq0$. Then the gauge degrees of freedom do not decouple and it
becomes quite unclear whether properties of the continuum theory (such
as unitarity) are reproduced in the continuum limit, after the effect
of dynamical gauge fields is taken into account. Basically this is the
origin of difficulties of chiral gauge theories on the lattice~[\SHA].
It is thus quite important to understand the structure of breakings of
the gauge symmetry on the lattice, which will be denoted
by~${\cal A}$, while keeping the lattice spacing {\it finite}.

What is the possible structure of~${\cal A}$ for~$a\neq0$? This
question appears meaningless unless certain conditions are imposed
on~${\cal A}$. After all, uniqueness of the gauge anomaly in the
continuum theory~[\BAR--\GRO,\BEC--\DUB] is lost for a finite
ultraviolet cutoff, and the explicit form of the breaking~${\cal A}$
is expected to depend strongly on the details of the lattice
formulation. But what kind of conditions can strongly constrain the
structure of~${\cal A}$? And, under such conditions, is it possible to
relate~${\cal A}$ and the anomaly in the continuum theory? It seemed
almost impossible to answer these questions. (This statement is not
completely true: If one restricts operators with the mass
dimension~$\leq5$ (we assign one mass dimension to the ghost field),
the complete classification of possible breakings has been known in
the context of the Rome approach~[\BOR].)

The atmosphere has changed after L\"uscher's theorem on the
$\gamma_5$-anomaly in the abelian lattice gauge theory~$G=U(1)$
appeared~[\LUS]. Assuming smoothness, locality\foot{The meaning of the
locality is of course different from that of the continuum theory. We
will explain this terminology in detail in the next section.} and the
topological nature of the anomaly, he proved the theorem for a
4-dimensional infinite lattice, which corresponds to\foot{For our
notation, see appendix~A.}
$$
\eqalign{
   {\cal A}&=\delta_B\ln\Det M'[A]
\cr
   &=\sum_nc(n)
   \Bigl[\alpha+\beta_{\mu\nu}F_{\mu\nu}(n)+
   \gamma\varepsilon_{\mu\nu\rho\sigma}F_{\mu\nu}(n)
   F_{\rho\sigma}(n+\widehat\mu+\widehat\nu)+\Delta_\mu^*k_\mu(n)
   \Bigr],
\cr
}
\eqn\onexone
$$
where $\Det M'$~is a fermion determinant and $\delta_B$ is the BRS
transformation~[\BEC] corresponding to the gauge transformation in the
abelian lattice gauge theory, $\delta_BA_\mu(n)=\Delta_\mu c(n)$
and~$\delta_Bc(n)=0$; $c(n)$ stands for the abelian Faddeev-Popov
ghost field defined on the lattice. In~eq.~\onexone, $\alpha$,
$\beta_{\mu\nu}$ and~$\gamma$ are constants, and $k_\mu(n)$ in the
last term is a {\it local\/} and {\it gauge invariant\/} current. Note
that eq.~\onexone\ holds for finite lattice spacing and that the
structure is quite independent of the details of the formulation. In
this sense, this theorem provides a universal characterization of the
gauge anomaly in abelian lattice chiral gauge theory. Moreover, the
theorem asserts that the anomaly cancellation in the abelian lattice
theory is (almost) equivalent to that of the continuum theory: The
first two constants vanish, $\alpha=\beta_{\mu\nu}=0$, if the anomaly
is a pseudoscalar quantity. The term proportional to~$\gamma$ is
cancelled if $\sum_Re_R^3-\sum_Le_L^3=0$. Here $e_H$ stands for the
$U(1)$~charge, because we have absorbed the $U(1)$ charge in~$c$ and
in~$F_{\mu\nu}$. Finally, the last term of the breaking~\onexone\ can
be removed by adding the {\it local\/}
counterterm~${\cal B}=\sum_nA_\mu(n)k_\mu(n)$ to the effective action
$\ln\Det M'\to\ln\Det M'+{\cal B}$, because $\delta_B{\cal B}=%
\sum_n\Delta_\mu c(n)k_\mu(n)=-\sum_nc(n)\Delta_\mu^*k_\mu(n)$. This
argument shows that the effective action with finite lattice spacing
can be made gauge invariant if (and only if) the fermion multiplet is
anomaly-free! This remarkable observation was fully utilized in the
existence proof of an exactly gauge invariant lattice formulation of
anomaly-free abelian chiral gauge theories~[\LUSC].

In this paper, we attempt to generalize the above
theorem~\onexone\ for general (compact) gauge groups. Our scheme is
somewhat different from that of~refs.~[\LUS,\LUSCH]. In~ref.~[\LUSCH],
this problem in nonabelian theories was shown to be equivalent to a
classification of gauge invariant topological fields in
$(4{+}2)$-dimensional space, where 4~dimensions are discrete and
2~dimensions are continuous. Instead, in this paper, we analyze
general nontrivial local solutions to the Wess-Zumino consistency
condition~[\WES] in lattice gauge theory. For a generic gauge group,
the BRS transformation is defined by:\foot{This transformation is
obtained by parameterizing the gauge transformation parameter
in~$U(n,\mu)\to g(n)^{-1}\*U(n,\mu)\*g(n+\widehat\mu)$
by~$g=\exp(\lambda c)$ where $\lambda$~stands for an infinitesimal
Grassmann parameter.}
$$
   \delta_BU(n,\mu)=U(n,\mu)c(n+\widehat\mu)-c(n)U(n,\mu),\qquad
   \delta_Bc(n)=-c(n)^2.
\eqn\onextwo
$$
Since this BRS transformation is nilpotent~$\delta_B^2=0$, the
breaking~${\cal A}=\delta_B\ln\Det M'$ must satisfy the Wess-Zumino
consistency condition
$$
   \delta_B{\cal A}=0,
\eqn\onexthree
$$
like in the continuum theory~[\WES,\BEC]. In the continuum theory,
consistency and uniqueness of anomaly-free chiral gauge theories on
the perturbative level follow from detailed analyses of the
consistency condition~[\BEC--\DUB] (for a more complete list of
references, see ref.~[\DUB]). We will see below that the consistency
condition~\onexthree, combined with the locality in the sense
of~ref.~[\LUS], strongly constrains the possible structure
of~${\cal A}$, as it does in the continuum theory. Our basic strategy
is to imitate as much as possible the procedure in the continuum
theory, especially that of~ref.~[\BRA]. Of course, there are many
crucial differences between continuum and lattice theories and how to
handle these differences becomes the key to our ``algebraic''
approach.

The organization of this paper is as follows. Our main theorems which
generalize eq.~\onexone\ are stated in section~3. Our theorems are
applicable only if the gauge anomaly~${\cal A}$ depends locally on
the gauge field. The only framework known at present, which possesses
this property is the formulation of~refs.~[\LUSC,\LUSCH] based on the
Ginsparg-Wilson Dirac operator~[\GIN--\NEU], or equivalently the
overlap formulation~[\NAR,\RAN]. Therefore, in section~2, we summarize
basic properties of the gauge anomaly along the formulation
of~ref.~[\LUSCH]. At the same time, we introduce notions of
admissibility and of locality. We also introduce the gauge potential
and define the so-called ``perturbative configuration.'' Sections~4 to
6 are entirely devoted to the determination of general nontrivial
local solutions to the consistency condition in the abelian
theory~$G=U(1)^N$. In~section~4, we give some preliminaries.
In~section~5, we prove several lemmas concerning cohomology on an
infinite lattice. Here the technique of noncommutative differential
calculus~[\CON--\DIN] turns out to be a powerful tool~[\FUJ,\FUJI].
Utilizing these lemmas, in section~6, we first determine a complete
list of nontrivial local solutions to the consistency condition in the
abelian theory. Here the ghost number of the solution is arbitrary.
Then we restrict the ghost number of the solution to unity. After
imposing several conditions, we obtain the content of the theorem for
the abelian theory. Section~7~is devoted to the nonabelian extension.
In section~7.1, we derive a basic lemma which guarantees the adjoint
invariance of nontrivial solutions. In~section~7.2, by using several
assumptions, we show the uniqueness of the nontrivial local anomaly to
all orders in powers of the gauge potential. This establishes the
content of our theorem for nonabelian theories, which will be stated
in section~3. In section~7.3, we explicitly write down such a
nontrivial local anomaly by utilizing the interpolation technique of
lattice fields~[\GOC,\HER]. The last section is devoted to concluding
remarks. Our notation is summarized in~appendix~A. In~appendix~B, we
explain the calculation of the Wilson line which appears in the
integrability condition of~ref.~[\LUSCH].

\chapter{Gauge anomaly in the Ginsparg-Wilson approach}
\section{Admissibility, locality and the gauge potential}
The ``admissible'' gauge field is defined by~[\LUSCH]
$$
   \|P(n,\mu,\nu)-1\|<\epsilon,\qquad\hbox{for all $n$, $\mu$, $\nu$},
\eqn\twoxone
$$
where $P(n,\mu,\nu)$ is the plaquette variable in the representation
to which the Weyl fermion belongs and $\epsilon$~is a certain small
positive constant. In this expression, $\|{\cal O}\|$~is the operator
norm defined by~[\REE]
$$
   \|{\cal O}\|=\sup_{v\neq0}{\|{\cal O}v\|\over\|v\|},
\eqn\twoxtwo
$$
where the norm on the right hand side is defined by the standard norm
for vectors. The reason for this restriction of field space is
two-fold:

Consider a finite lattice. Let us suppose that the Dirac operator
satisfies an index theorem. Namely, a difference of numbers of
normalizable zero modes of the Dirac operator with opposite chirality
is equal to the topological charge of the gauge field configuration.
The index is an integer and thus inevitably jumps even if the gauge
field configuration changes smoothly. This argument suggests that such
a Dirac operator cannot be a smooth function of the gauge field.
Smoothness of the Dirac operator and in turn that of the gauge anomaly
are thus expected to hold only within a restricted field space. In
fact, a detailed analysis~[\HERN] of Neuberger's overlap Dirac
operator~[\NEU], which satisfies the index theorem~[\HASE,\LUSCHE],
shows that the Dirac operator depends smoothly and locally on the
gauge field when~$\epsilon\leq1/30$ in~eq.~\twoxone. Our proof is
valid only when the gauge anomaly depends on the gauge field smoothly
and locally.

Closely related to the above point, we note that any configuration of
the lattice gauge field can smoothly be deformed into the trivial one,
$U(n,\mu)=1$, and thus the topology of the gauge field space is
trivial if no restriction is imposed. On the other hand, it has been
known~[\LUSCHER] that, under the condition~\twoxone, one can define a
nontrivial principal bundle over a periodic lattice such that the
field space is divided into topological sectors. For example, for the
fundamental representation of~$SU(2)$, $\epsilon\leq0.015$ is enough
for the construction of~ref.~[\LUSCHER] to work. Later we will utilize
the interpolation method of~ref.~[\GOC] which is based on the section
of the principal bundle of~ref.~[\LUSCHER].

Note that eq.~\twoxone\ is a gauge invariant condition. The gauge
transformed configurations of an admissible configuration are all
admissible. However, the structure of the space of admissible
configurations is quite complicated, and no simple parameterization in
terms of the gauge potential has been known except for abelian
cases~[\LUS]. This is the reason why our theorem for nonabelian
theories is in practice applicable only for the ``perturbative
configurations'' which will be explained below.

As noted in the introduction, our basic strategy is to imitate the
argument in the continuum theory. The first important difference from
the continuum theory is the notion of locality. The anomaly is a
local quantity when the ultraviolet cutoff is sent to infinity. But of
course this is not the case for~$a\neq0$ so that we need an
appropriate notion which works on the lattice. Here we follow the
definition of~ref.~[\LUS] (see also~ref.~[\NIED]). Suppose that
$\phi(n)$~is a field on the lattice which depends on link
variables~$U$. The field~$\phi(n)$ may depend on the link
variable~$U(m,\mu)$ at a distant link from the site~$n$. We say that
$\phi(n)$ locally depends on the link variable, if this dependence
on~$U(m,\mu)$ becomes exponentially weak as~$|n-m|\to\infty$. To be
more precise, consider the following decomposition:
$$
   \phi(n)=\sum_{k=1}^\infty\phi_k(n),
\eqn\twoxthree
$$
where $\phi_k(n)$~depends only on link variables~$U$ inside a block of
size~$k$ centered at the site~$n$ (such a field~$\phi_k(n)$ is called
ultra-local). If all these fields~$\phi_k(n)$ and their
derivatives~$\phi_k(n;m_1,\mu_1;\cdots;m_N,\mu_N)$ with respect to
the link variables $U(m_1,\mu_1)$, \dots, $U(m_N,\mu_N)$ are bounded
as
$$
   \bigl|\phi_k(n;m_1,\mu_1;\cdots;m_N,\mu_N)\bigr|
   \leq C_Nk^{p_N}\exp(-\theta k),
\eqn\twoxfour
$$
by the constants $C_N$, $p_N$ and $\theta$, which are all independent
of link variable configurations, then we say that $\phi(n)$ locally
depends on the link variable. In what follows, we also introduce the
gauge potential and the ghost field. The same terminology will be used
by simply replacing ``link variable'' by the name of each field. When
no confusion arises, we say simply that $\phi(n)$~is local. Also when
a functional is given by a sum of such local fields,
$\Phi=\sum_n\phi(n)$, we simply say that $\Phi$ is local. If
$\phi(n)$~is a local field, the effective range of dependences is a
finite number in lattice units. Physically, therefore, this locality
can be regarded as equivalent to ultra-locality. The technical reason
for this definition of locality is that the Dirac operator which
satisfies the Ginsparg-Wilson relation cannot be ultra-local in
general~[\HOR,\BIE] and, on the other hand, we can apply the
Poincar\'e lemma of ref.~[\LUS] if dependences are exponentially
suppressed.

The basic degrees of freedom in lattice gauge theory are link
variables. However we prefer to stick to the gauge potential, because
its use is essential for arguments in the continuum theory. To stretch
the validity of our argument as far as possible, we consider the
following two cases.

\noindent
Case~I. When the gauge group is abelian~$G=U(1)^N$. If we take
$0<\epsilon\leq1$ in~eq.~\twoxone\ or equivalently (the
superscript~$a$ here labels each $U(1)$ factor in~$G$)
$$
   \|\Ln P^a(n,\mu,\nu)\|<{\pi\over3},\qquad
   \hbox{for all $a$, $n$, $\mu$, $\nu$},
\eqn\twoxfive
$$
there exists a relatively simple prescription~[\LUS] which allows a
complete parameterization of the space of admissible gauge fields.
Under the condition~\twoxfive, one can associate the abelian gauge
potential to the link variable such that
$$
   U^a(n,\mu)=\exp A_\mu^a(n),\qquad
   -\infty<{1\over i}\,A_\mu^a(n)<\infty,
\eqn\twoxsix
$$
and moreover
$$
   \Ln P^a(n,\mu,\nu)=\Delta_\mu A_\nu^a(n)-\Delta_\nu A_\mu^a(n).
\eqn\twoxseven
$$
{}From this relation and~eq.~\twoxfive, one concludes that if a
configuration~$A_\mu^a$ is admissible, the rescaled one~$tA_\mu^a$
with~$0\leq t\leq1$ is also admissible. In this prescription~[\LUS]
(a closely related prescription for 2-dimensional periodic lattices
was first given in~ref.~[\HER]), the abelian gauge
potential~$A_\mu^a(n)$ corresponding to the given link
variables~$U^a(n,\mu)$ is not unique. Also this mapping does not
preserve the locality. Nevertheless, as far as gauge invariant
quantities are concerned, such an ambiguity disappears and the
locality becomes common for both variables. See~refs.~[\LUS,\FUJIW]
for details.

\noindent
Case~II. For a general (compact) gauge group~$G$, we define
$$
   U(n,\mu)=\exp A_\mu(n),\qquad\|A_\mu(n)\|\leq\pi,
\eqn\twoxeight
$$
and we further impose
$$
   \|A_\mu(n)\|<{1\over4}\ln(1+\epsilon)\leq\pi,\qquad
   \hbox{for all $\mu$ and $n$}.
\eqn\twoxnine
$$
By noting $\|{\cal O}+{\cal O}'\|\leq\|{\cal O}\|+\|{\cal O}'\|$ and
$\|{\cal OO}'\|\leq\|{\cal O}\|\|{\cal O}'\|$~[\REE], one can see that
configurations which satisfy eq.~\twoxnine\ are in fact admissible,
i.e., they satisfy eq.~\twoxone. Note that eq.~\twoxnine\ is a very
restrictive condition and contains only a small portion of admissible
configurations; in fact, the condition~\twoxnine\ is not gauge
invariant. We call configurations which
satisfy~eq.~\twoxnine\ ``perturbative.'' For perturbative
configurations, all link variables are close to unity
$\|U(n,\mu)-1\|<\epsilon'=(1+\epsilon)^{1/4}-1$. Unfortunately, our
theorem for nonabelian theories is applicable only for this restricted
space, when the admissibility~\twoxone\ is required.

\section{Fermion determinant and the gauge anomaly}
In this subsection, we study basic properties of the gauge anomaly
appearing in the formulation based on the Ginsparg-Wilson Dirac
operator~[\LUSC,\LUSCH] with a {\it particular\/} choice of the
integration measure. As noted sometimes~[\AOY,\NEUB], this formulation
can be reinterpreted in terms of the overlap formulation~[\NAR,\RAN].
Therefore it must be possible to repeat a similar argument also in the
context of the overlap formulation.

Following refs.~[\LUSC,\LUSCH], we define the fermion determinant as
$$
   \Det M'=\int d[\psi]d[\overline\psi]\,
   \exp\biggl[-\sum_n\overline\psi(n)D\psi(n)\biggr],
   \qquad\widehat P_H\psi(n)=\psi(n),
\eqn\twoxten
$$
where the Dirac operator~$D$ satisfies the Ginsparg-Wilson
relation~$\gamma_5D+D\gamma_5=D\gamma_5D$~[\GIN]. We assume that the
Dirac operator~$D$ is gauge covariant and local in the sense
of~ref.~[\LUSC] and depends smoothly on the gauge field. Thus we
assume the admissibility~\twoxone\ for gauge field configurations. The
chirality of the fermion is defined with respect to the
Ginsparg-Wilson chiral matrix~$\widehat\gamma_5=%
\gamma_5(1-D)$~[\NARA,\NIED,\KIK]. Namely, the projection operator
has been defined by~$\widehat P_H=(1+\epsilon_H\widehat\gamma_5)/2$.
The chirality of the anti-fermion is, on the other hand, defined by
the conventional~$\gamma_5$ matrix.

The integration measure for the fermion~$d[\psi]$
in~eq.~\twoxten\ thus depends on the gauge field nontrivially due to
the condition $\widehat P_H\psi=\psi$. However this condition alone
does not specify the integration measure uniquely. For definiteness,
we make the following choice which starts with the particular
``measure term''~[\SUZ]\foot{${\cal L}_\eta'$ identically vanishes
when the representation of the Weyl fermion is (pseudo-)real~[\SUZ].}
$$
   {\cal L}_\eta'=-i\epsilon_H\int\nolimits_0^1ds\,
   \Tr\widehat P_H
   \bigl[\partial_s\widehat P_H,\delta_\eta\widehat P_H\bigr],
\eqn\twoxeleven
$$
where $\Tr$~stands for the summation over lattice points~$\sum_n$ of
the diagonal~$(n,n)$ components as well as traces over the gauge and
the spinor indices. In the above expression, $\eta$~stands for the
infinitesimal variation of link variables
$$
   \delta_\eta U(n,\mu)=\eta_\mu(n)U(n,\mu).
\eqn\twoxtwelve
$$
We have to specify the $s$-dependence in~eq.~\twoxeleven. As a simple
choice, we take
$$
   U(n,s,\mu)=\exp\bigl[sA_\mu(n)\bigr],\qquad 0\leq s\leq1,
\eqn\twoxthirteen
$$
for both cases I~\twoxsix\ and~II~\twoxeight\ above. Note that the
line in the configuration space~$U(n,s,\mu)$ which connects 1
and~$U(n,\mu)$ is contained in the admissible space~\twoxone\ and, for
the case~II, in the perturbative region~\twoxnine. The
functional~\twoxeleven\ depends smoothly and locally on the gauge
potential due to the assumed properties of the Dirac operator
(${\cal L}_\eta'$~does not contain the inverse of the Dirac operator).
Since the functional~${\cal L}_\eta'$ is linear in~$\eta_\mu$, it may
be written as
$$
   {\cal L}_\eta'=\sum_n\eta_\mu^a(n)j_\mu^{\prime a}(n),
   \qquad\eta_\mu(n)=\eta_\mu^a(n)T^a.
\eqn\twoxfourteen
$$
This current $j_\mu^{\prime a}$ depends smoothly and locally on the
gauge potential.

Now, using the Ginsparg-Wilson relation, one can show~[\SUZ] that
${\cal L}_\eta'$~satisfies the differential form of the integrability
condition~[\LUSC,\LUSCH]:\foot{Here we assume that $\eta$ and~$\zeta$
are independent of the gauge field.} 
$$
   \delta_\eta{\cal L}_\zeta'-\delta_\zeta{\cal L}_\eta'
   +{\cal L}_{[\eta,\zeta]}'
   =-i\epsilon_H\Tr\widehat P_H
   \bigl[\delta_\eta\widehat P_H,\delta_\zeta\widehat P_H\bigr].
\eqn\twoxfifteen
$$
Moreover, by considering a one-parameter family of gauge fields,
$U_t(n,\mu)$ ($0\leq t\leq1$) and introducing the transporting
operator~$Q_t$ by~[\LUSCH]
$$
   \partial_tQ_t=[\partial_tP_t,P_t]Q_t,
   \qquad P_t=\widehat P_H|_{U\to U_t},\qquad Q_0=1,
\eqn\twoxsixteen
$$
one can show (appendix~B) that ${\cal L}_\eta'$ satisfies the
integrability in the integrated form~[\LUSCH] for an arbitrary closed
{\it loop\/} $U_0(n,\mu)=U_1(n,\mu)$ (here $\eta_\mu(n)=%
\partial_tU_t(n,\mu)U_t(n,\mu)^{-1}$)
$$
   W'=\exp\biggl(i\int\nolimits_0^1dt\,{\cal L}_\eta'\biggr)
   =\Det(1-P_0+P_0Q_1)^{-\epsilon_H},
\eqn\twoxseventeen
$$
as long as the loop $U_t(n,\mu)$ is contained within the perturbative
region~\twoxnine\ for the case~II. Since both the space of admissible
fields~\twoxfive\ for case~I and the space of perturbative
configurations~\twoxnine\ for case~II are contractable, there is no
global obstruction~[\BAER] which is a lattice counterpart of the
Witten's anomaly~[\WIT]. Eq.~\twoxseventeen\ guarantees that there
exists an integration measure~$d[\psi]d[\overline\psi]$ which
corresponds to the measure term~${\cal L}_\eta'$~[\LUSCH]. In
particular, the infinitesimal variation of the fermion
determinant~\twoxten\ is given by
$$
   \delta_\eta\ln\Det M'=\Tr\delta_\eta D\widehat P_HD^{-1}
   +i\epsilon_H{\cal L}_\eta'.
\eqn\twoxeighteen
$$

We have completely specified the fermion determinant~\twoxten\ up to
a physically irrelevant proportionality constant. This fermion
determinant is, however, not gauge invariant in general. The resulting
gauge anomaly~${\cal A}=\delta_B\ln\Det M'$ is obtained simply by
setting
$$
   \eta_\mu(n)=U(n,\mu)c(n+\widehat\mu)U(n,\mu)^{-1}-c(n)
\eqn\twoxnineteen
$$
in~eq.~\twoxtwelve. Then from eqs.~\twoxeighteen\ and~\twoxfourteen,
we have
$$
   {\cal A}=\epsilon_H\Tr c\gamma_5\biggl(1-{1\over2}D\biggr)
   -i\epsilon_H\sum_nc^a(n)
   \Bigl[j_\mu'(n)-U(n-\widehat\mu,\mu)^{-1}j_\mu'(n-\widehat\mu)
         U(n-\widehat\mu,\mu)\Bigr]^a,
\eqn\twoxtwenty
$$
where use of the gauge covariance~$\delta_BD=[D,c]$ for~$s=1$ and the
Ginsparg-Wilson relation has been made. Manifestly, this
anomaly~${\cal A}$ depends smoothly and locally on the gauge potential
(and on the ghost field) from the assumed properties of the Dirac
operator and from the properties of the current~$j_\mu^{\prime a}$.

Let us next study the classical continuum limit of~${\cal A}$. The
gauge potential in the classical continuum limit~$A_\mu(x)$ is
introduced by the conventional manner
$$
   U(n,\mu)={\cal P}\exp\biggl[
   a\int\nolimits_0^1du\,A_\mu(n+(1-u)\widehat\mu a)\biggr],
\eqn\twoxtwentyone
$$
where ${\cal P}$ stands for the path-ordered product. Then the first
term of~eq.~\twoxtwenty\ produces the covariant gauge anomaly which
can be deduced from the general arguments~[\FUJIK,\LUSCH] or from
explicit calculations using Neuberger's overlap Dirac
operator~[\KIKU--\SUZU] as\foot{Of course, we assume that parameters
in the Dirac operator has been chosen such that there is only one
massless degree of freedom.} 
$$
   \epsilon_H\Tr c\gamma_5\biggl(1-{1\over2}D\biggr)
   \buildrel{a\to0}\over\rightarrow
   -{\epsilon_H\over8\pi^2}\int d^4x\,\varepsilon_{\mu\nu\rho\sigma}
   \tr c\,\partial_\mu\biggl(A_\nu\partial_\rho A_\sigma
   +{2\over3}A_\nu A_\rho A_\sigma\biggr),
\eqn\twoxtwentytwo
$$
for a single Weyl fermion. For the second term of~eq.~\twoxtwenty,
which corresponds to a divergence of the so-called Bardeen-Zumino
current~[\BARD] in the continuum theory, it is easier to
consider~${\cal L}_\eta'$~\twoxeleven\ instead of the divergence of
the current~$j_\mu^{\prime a}$. With the choice~\twoxthirteen, we see
in the classical continuum limit\foot{For {\it abelian\/} cases, the
relation $\delta_B\widehat P_H=s[\widehat P_H,c]$ holds for
arbitrary~$a$.}
$$
   \delta_B\widehat P_H=s[\widehat P_H,c]+O(a).
\eqn\twoxtwentythree
$$
Then by using the Ginsparg-Wilson relation, we have
$$
   i\epsilon_H{\cal L}_\eta'
   =-{\epsilon_H\over2}\int\nolimits_0^1ds\,s\,\partial_s\Tr
   c\gamma_5(1-D)+O(a).
\eqn\twoxtwentyfour
$$
It is possible to argue that the $O(a)$-term
in~eq.~\twoxtwentythree\ contributes only to the $O(a)$-term
in~eq.~\twoxtwentyfour\ from the mass dimension and the pseudoscalar
nature of~$i\epsilon_H{\cal L}_\eta'$ (assuming that Lorentz
invariance is restored in the classical continuum
limit).\foot{Strictly speaking, an explicit calculation
of~eq.~\twoxeleven\ or of~eq.~\twoxtwentyfour\ in the classical
continuum limit, using, say, Neuberger's overlap Dirac operator, has
not been carried out in the literature. A corresponding calculation on
the linearized level in the overlap formulation was given in the last
reference of~ref.~[\RAN]. See also~ref.~[\LUSCH]} Then,
{}from~eq.~\twoxtwentytwo\ with the substitution~$A_\mu\to sA_\mu$, we
have
$$
   i\epsilon_H{\cal L}_\eta'
   \buildrel{a\to0}\over\rightarrow
   {\epsilon_H\over8\pi^2}\int d^4x\,\varepsilon_{\mu\nu\rho\sigma}
   \tr c\,\partial_\mu\biggl({2\over3}A_\nu\partial_\rho A_\sigma
   +{1\over2}A_\nu A_\rho A_\sigma\biggr).
\eqn\twoxtwentyfive
$$
Combining eqs.~\twoxtwentytwo\ and~\twoxtwentyfive, we have the
correct consistent anomaly in the continuum theory:
$$
   {\cal A}\buildrel{a\to0}\over\rightarrow
   -{\epsilon_H\over24\pi^2}\int d^4x\,\varepsilon_{\mu\nu\rho\sigma}
   \tr c\,\partial_\mu\biggl(A_\nu\partial_\rho A_\sigma
   +{1\over2}A_\nu A_\rho A_\sigma\biggr).
\eqn\twoxtwentysix
$$
This expression is for a simple gauge group. The gauge anomaly for a
generic gauge group~$G=\prod_\alpha G_\alpha$ can be obtained by
simply substituting~$c\to\sum_\alpha c^{G_\alpha}$
and~$A_\mu\to\sum_\alpha A_\mu^{G_\alpha}$ in~eq.~\twoxtwentysix. To
have the standard form of the anomaly for~$G=\prod_\alpha G_\alpha$,
we add the local counterterm to the effective action, $\ln\Det M''=%
\ln\Det M'+{\cal S}$, or to the measure term~$i\epsilon_H{\cal L}''=%
i\epsilon_H{\cal L}'+\delta_\eta{\cal S}$, where
$$
   {\cal S}={\epsilon_H\over144\pi^2}
   \sum_n\varepsilon_{\mu\nu\rho\sigma}
   V(n,\mu)^{U(1)_\beta}\tr V(n,\nu)^{(\alpha)}V(n,\rho)^{(\alpha)}
   V(n,\sigma)^{(\alpha)},
\eqn\twoxtwentyseven
$$
and $V(n,\mu)=\bigl[U(n,\mu)-U(n,\mu)^\dagger\bigr]/2$.
The superscript~$\alpha$ runs over simple groups in~$G$, and
$\beta$~denotes each $U(1)$~factor in~$G$. ${\cal S}$~depends smoothly
and locally on the link variable and the modification does not affect
the integrability, eqs.~\twoxfifteen\ and~\twoxseventeen.\foot{Note
that (provided that $\eta$ and~$\zeta$ are independent of the gauge
field) the relation $(\delta_\eta\delta_\zeta-%
\delta_\zeta\delta_\eta+\delta_{[\eta,\zeta]}){\cal S}=0$ holds for
any functional~${\cal S}$ of the link variable.} The
counterterm~${\cal S}$ was chosen such that its classical continuum
limit becomes $\epsilon_H\int d^4x\,\varepsilon_{\mu\nu\rho\sigma}%
A_\mu^{U(1)_\beta}\tr A_\nu^{(\alpha)}A_\rho^{(\alpha)}%
A_\sigma^{(\alpha)}/(144\pi^2)$. Then the gauge anomaly of the
modified effective action becomes
$$
\eqalign{
   {\cal A}&\buildrel{a\to0}\over\rightarrow
   -{\epsilon_H\over24\pi^2}\int d^4x\,
   \biggl\{
   \varepsilon_{\mu\nu\rho\sigma}\tr c^{(\alpha)}
   \partial_\mu\biggl[A_\nu^{(\alpha)}
   \partial_\rho A_\sigma^{(\alpha)}
   +{1\over2}A_\nu^{(\alpha)}
   A_\rho^{(\alpha)}A_\sigma^{(\alpha)}\biggr]
\cr
   &\qquad\qquad\qquad\qquad\quad
   +\varepsilon_{\mu\nu\rho\sigma}c^{U(1)_\beta}
   \partial_\mu A_\nu^{U(1)_\beta}
   \partial_\rho A_\sigma^{U(1)_\beta}
\cr
   &\qquad\qquad\qquad\qquad\quad
   +\varepsilon_{\mu\nu\rho\sigma}c^{U(1)_\beta}
   \tr\partial_\mu\biggl[A_\nu^{(\alpha)}
   \partial_\rho A_\sigma^{(\alpha)}
   +{2\over3}A_\nu^{(\alpha)}
   A_\rho^{(\alpha)}A_\sigma^{(\alpha)}\biggr]
\cr
   &\qquad\qquad\qquad\qquad\quad
   +2\varepsilon_{\mu\nu\rho\sigma}\tr\Bigl[c^{(\alpha)}
   \partial_\mu A_\nu^{(\alpha)}\Bigr]
   \partial_\rho A_\sigma^{U(1)_\beta}\biggr\},
\cr
}
\eqn\twoxtwentyeight
$$
in the classical continuum limit.

Finally, we have to mention a topological property of the gauge
anomaly which is associated with each $U(1)$~factor. Going back to the
infinitesimal variation~\twoxeighteen\ for~$a\neq0$, the gauge anomaly
is obtained by substituing $\delta_\eta\to\delta_B$. Since, as already
noted,
$$
   \delta_B\widehat P_H=s[\widehat P_H,c^{U(1)_\beta}]
   +(\hbox{terms proportional to $c^{(\alpha)}$}),
\eqn\addone
$$
where $\alpha$ stands for simple groups, we have~[\SUZ] after using
the Ginsparg-Wilson relation
$$
   {\cal A}=\epsilon_H\int_0^1ds\,\Tr c^{U(1)_\beta}
   \gamma_5\biggl(1-{1\over2}D\biggr)
   +(\hbox{terms proportional to $c^{(\alpha)}$}).
\eqn\addtwo
$$
The combination~$q=\tr\gamma_5(1-D/2)$ appearing here is a topological
field~[\HASE,\LUSCHE] such that
$$
   \sum_n\delta q(n)=0,
\eqn\addthree
$$
for an arbitrary local variation of the gauge field~$\delta$.
Therefore we see that
$$
   \delta{\cal A}=(\hbox{terms proportional to $c^{(\alpha)}$}),
   \qquad{\rm for}\quad c^{U(1)_\beta}(n)\to{\rm const.},
\eqn\addfour
$$
where $\delta$~is an arbitrary local variation of the gauge potential.
Note that, since $\delta_BA_\mu\to%
(\hbox{terms prop.\ to $c^{(\alpha)}$})$
for~$c^{U(1)_\beta}(n)\to{\rm const.}$, this topological property for
abelian factors holds even after the addition of {\it local\/} terms
to the effective action, such as~${\cal S}$ in~eq.~\twoxtwentyseven.

We have thus observed that (I)~the anomaly on the lattice~${\cal A}=%
\delta_B\ln\Det M''$ depends smoothly and locally on the gauge
potential and on the ghost field, (II)~${\cal A}$ reproduces the
gauge anomaly in the continuum theory in the classical continuum
limit as in~eq.~\twoxtwentyeight, and (III)~$U(1)$~gauge anomalies
in~${\cal A}$ have the topological property~\addfour. In the following
sections, we show that such an anomaly~${\cal A}=\delta_B\ln\Det M''$
on an infinite lattice can always be written as~${\cal A}=%
\delta_B{\cal B}$, where ${\cal B}$~depends smoothly and
{\it locally\/} on the gauge potential, if (and only if) the anomaly
in the continuum theory is canceled, $\tr_{R-L}T^a\{T^b,T^c\}=0$~etc.
This statement holds to all orders in powers of the gauge potential
(for nonabelian cases). This implies that, for an anomaly-free fermion
multiplet, one can improve the fermion determinant according to
$$
   \Det M''[A]\to\Det M[A]=\Det M''[A]\exp(-{\cal B}[A]),
\eqn\twoxtwentynine
$$
so that the improved fermion determinant~$\Det M[A]$ has the exact
gauge invariance.\foot{Since the fermion determinant~$\Det M[A]$ is
gauge invariant, one can then regard it as a functional of the link
variable~$\Det M[U]$ for the case~I (this is trivially the case for
the case~II).} Therefore, to all orders of the gauge potential, there
exists a gauge invariant lattice formulation of anomaly-free
nonabelian chiral gauge theories, as far as the perturbative
configurations~\twoxnine\ on an infinite lattice are concerned.

In the context of the overlap formulation, our choice of the measure
term~\twoxeleven\ corresponds to a particular choice of the phase of
the vacuum state. The formula corresponding to~eq.~\twoxtwenty\ was
given in the last reference of~ref.~[\RAN]. The gauge anomaly and
Witten's anomaly as local and global obstructions in the overlap
formulation were studied in detail in~ref.~[\NEUBE]. See also
ref.~[\NEUB].

\chapter{Results}
In this section, we present our main results in a summarized form.
For the abelian gauge group~$G=U(1)^N$, we will show the following
theorem.
\proclaim Theorem (Abelian theory).
Let ${\cal A}[c,A]$ be the gauge anomaly defined on a 4-dimensional
infinite hypercubic lattice. Suppose that (I)~${\cal A}$~depends
smoothly and locally on the abelian gauge potential~$A_\mu^a$ and on
the abelian ghost field~$c^a$, (II)~${\cal A}$ reproduces for smooth
field configurations the gauge anomaly in the continuum
theory~\twoxtwentyeight\ in the classical continuum limit, and
(III)~${\cal A}$ has the topological property
$$
   \delta{\cal A}=0,
   \qquad{\rm for}\quad c^a(n)\to{\rm const.},
\eqn\addfive
$$
where $\delta$~is an arbitrary local variation of the gauge potential.
Then ${\cal A}$ is always of the form (for a single Weyl fermion)
$$
   {\cal A}[c,A]=-{\epsilon_H\over96\pi^2}\sum_n
   \sum_{abc}\varepsilon_{\mu\nu\rho\sigma}c^a(n)
   F_{\mu\nu}^b(n)F_{\rho\sigma}^c(n+\widehat\mu+\widehat\nu)
   +\delta_B{\cal B}[A],
\eqn\threexone
$$
where the functional ${\cal B}$ depends smoothly and locally on the
gauge potential~$A_\mu^a$.

This is a natural generalization of L\"uscher's result~\onexone\ to
multi-$U(1)$~cases. Thus, under the prerequisites of the theorem, the
anomaly cancellation in the corresponding continuum theory
$\sum_Re_R^ae_R^be_R^c-\sum_Le_L^ae_L^be_L^c=0$ guarantees the gauge
invariance of the effective action, after subtracting the local
counterterm~${\cal B}$.

For nonabelian gauge theories, we will show the following statement.
\proclaim Theorem (Nonabelian theory).
Let ${\cal A}[c,A]$ be the gauge anomaly defined on a 4-dimensional
infinite hypercubic lattice. Suppose that (I)~${\cal A}$~depends
smoothly and locally on the gauge potential~$A_\mu$ and on the ghost
field~$c$, (II)~${\cal A}$ reproduces for smooth field configurations
the gauge anomaly in the continuum theory~\twoxtwentyeight\ in the
classical continuum limit, and (III)~$U(1)$ gauge anomalies
in~${\cal A}$ have the topological property
$$
   \delta{\cal A}=(\hbox{terms proportional to $c^{(\alpha)}$}),
   \qquad{\rm for}\quad c^{U(1)_\beta}(n)\to{\rm const.},
\eqn\addsix
$$
where $\delta$~is an arbitrary local variation of the gauge potential.
Then if the anomaly in the corresponding continuum theory cancels,
$\tr_{R-L} T^a\{T^b,T^c\}=0$~etc., ${\cal A}$ is always BRS trivial,
i.e., ${\cal A}=\delta_B{\cal B}[A]$, where the functional~${\cal B}$
depends smoothly and locally on the gauge potential~$A_\mu$. This
statement holds to all orders in powers of the gauge
potential~$A_\mu$. The explicit form of the nontrivial
anomaly~${\cal A}\neq\delta_B{\cal B}$ is given in~eq.~(7.50).

Therefore, to all orders in powers of the gauge potential, the anomaly
cancellation in the continuum theory guarantees that of the lattice
theory. This seems remarkable but is not entirely unexpected. Let us
recall the expression in the classical continuum
limit~\twoxtwentyeight. In fact the expression holds to {\it all\/}
orders in powers of the lattice spacing~$a$ in the classical continuum
limit~$a\to0$ (we assume that the Lorentz covariance is restored in
this limit). In the classical continuum limit, each coefficient of the
expansion with respect to~$a$ is a local functional of the gauge
potential and the ghost field. Then the uniqueness theorem of
nontrivial anomalies in the continuum theory~[\BRA,\DUB] can be
invoked and one concludes that the anomaly~\twoxtwentyeight\ is the
unique possibility (up to contributions of local counterterms).
Therefore the anomaly cancellation~$\tr_{R-L}T^a\{T^b,T^c\}=0$
guarantees that ${\cal A}=\delta_B{\cal B}$ to {\it all\/} orders in
powers of the lattice spacing~$a$ in the classical continuum
limit~$a\to0$. Of course the expansion with respect to~$a$ is
(presumably at most) asymptotic and this does not prove the anomaly
cancellation for~$a\neq0$. Nevertheless, this argument makes the
content of the above theorem quite plausible. Finally, we emphasize
that the above theorems themselves do not assume the Ginsparg-Wilson
relation such that they are applicable to any formulation if the
prerequisites of the theorems are fulfilled.

\chapter{Preliminaries in the abelian theory}
\section{Noncommutative differential calculus}
To determine general nontrivial local solutions to the consistency
condition~\onexthree, we need cohomological information, as in the
continuum theory~[\BRA,\DUB]. To discuss $d$-cohomology on an infinite
lattice, the technique of noncommutative differential
calculus~[\CON--\DIN] is very useful, because it makes the standard
Leibniz rule of the exterior derivative valid even on the lattice. In
fact, this technique was applied successfully~[\FUJ,\FUJI] to an
algebraic proof of the higher dimensional extension of L\"uscher's
theorem of~ref.~[\LUS] (which is basically equivalent to
eq.~\onexone). Here we recapitulate its basic setup.

The bases of the 1-form on $D$-dimensional infinite hypercubic
lattice are defined as objects which satisfy the Grassmann algebra
$$
   dx_1,dx_2,\cdots,dx_D,\qquad dx_\mu dx_\nu=-dx_\nu dx_\mu.
\eqn\fourxone
$$
A generic $p$-form is defined by
$$
   f(n)={1\over p!}\,f_{\mu_1\cdots\mu_p}(n)
   \,dx_{\mu_1}\cdots dx_{\mu_p},
\eqn\fourxtwo
$$
where the summation over repeated indices is understood. The exterior
derivative is then defined by the forward difference operator as
$$
   df(n)={1\over p!}\,\Delta_\mu f_{\mu_1\cdots\mu_p}(n)
   \,dx_\mu dx_{\mu_1}\cdots dx_{\mu_p}.
\eqn\fourxthree
$$
The nilpotency of the exterior derivative~$d^2=0$ follows from this
definition. The essence of the noncommutative differential calculus
on infinite lattice is
$$
   dx_\mu f(n)=f(n+\widehat\mu)\,dx_\mu,
\eqn\fourxfour
$$
where $f(n)$~is a 0-form (i.e., a function). That is, a function on
the lattice and the basis of a 1-form do not simply commute. The
argument of the function is shifted along $\mu$-direction by one unit
when commuting these two objects. The remarkable fact, which follows
from the noncommutativity~\fourxfour, is that the standard Leibniz
rule of the exterior derivative~$d$ holds.
With~eqs.~\fourxthree\ and~\fourxfour, one can easily confirm that
$$
   d\,[f(n)g(n)]=df(n)g(n)+(-1)^pf(n)dg(n),
\eqn\fourxfive
$$
for forms $f(n)$ and~$g(n)$ where $f(n)$ is a $p$-form. The validity
of this Leibniz rule is quite helpful for following analyses.

We also introduce the abelian gauge potential 1-form and the abelian
field strength 2-form by
$$
   A^a(n)=A_\mu^a(n)\,dx_\mu,\qquad
   F^a(n)={1\over2}F_{\mu\nu}^a(n)\,dx_\mu dx_\nu=dA^a(n).
\eqn\fourxsix
$$
Note that the Bianchi identity takes the form~$dF^a(n)=0$. We will
{\it never\/} use the symbol~$F^a$ or~$F$ for {\it nonabelian\/}
field strength 2-form.

\section{Abelian BRS transformation}
Using~eqs.~\onextwo\ and~\twoxsix, the BRS transformation for the
gauge potential and for the ghost field in the abelian theory is given
by
$$
   \delta_BA_\mu^a(n)=\Delta_\mu c^a(n),\qquad
   \delta_Bc^a(n)=0.
\eqn\fourxseven
$$
The BRS transformation is nilpotent~$\delta_B^2=0$ and the abelian
field strength is BRS invariant~$\delta_BF_{\mu\nu}^a(n)=0$. We also
introduce the Grassmann coordinate~$\theta$~[\FER--\BONO] and define
the BRS exterior derivative by
$$
   s=\delta_B\,d\theta.
\eqn\fourxeight
$$
The usual 1-form~$dx_\mu$ and the BRS 1-form~$d\theta$ anticommute
with each other~$dx_\mu d\theta=-d\theta dx_\mu$ and the BRS
1-form~$d\theta$ commutes with itself $d\theta d\theta\neq0$.
Therefore, for a Grassmann-even (-odd) $p$-form~$f(n)$, we have
$$
   d\theta f(n)=\pm(-1)^pf(n)\,d\theta.
\eqn\fourxnine
$$
We also have
$$
   s^2=\{s,d\}=0,
\eqn\fourxten
$$
where the first relation follows from~$\delta_B^2=0$. Finally, we
introduce the ghost 1-form by
$$
   C^a(n)=c^a(n)\,d\theta.
\eqn\fourxeleven
$$
In terms of these forms, the BRS transformation in the abelian
theory~\fourxseven\ is expressed as
$$
   sA^a(n)=-dC^a(n),\qquad sC^a(n)=0,\qquad sF^a(n)=0.
\eqn\fourxtwelve
$$
The noncommutative rule~\fourxfour\ will always be assumed in
expressions written in terms of differential forms.

\chapter{Basic lemmas in the abelian theory}
Using the tools introduced in the preceding section, we establish in
this section several lemmas which provide cohomological information.
The algebraic Poincar\'e lemma specifies the $d$-cohomology on local
functions of the gauge potential and the ghost field. We use the
Poincar\'e lemma on an infinite lattice~[\LUS], which might be
regarded as a triviality of the de~Rham cohomology, to prove this
lemma. We next determine the BRS cohomology in the abelian
theory~$G=U(1)^N$. Finally the covariant Poincar\'e lemma tells the
$d$-cohomology on $s$-invariant functions. In the terminology
of~ref.~[\DUB], these three lemmas correspond to ${\rm H}(d)$,
${\rm H}(\delta)$ and~${\rm H}({\rm H}(\delta),d)$, respectively.
\section{Algebraic Poincar\'e lemma}
The algebraic Poincar\'e lemma\foot{The present algebraic Poincar\'e
lemma is somewhat different from that of~ref.~[\FUJ]. Practically, the
present form is more convenient.} asserts that a $d$-closed
form-valued local function on a $D$-dimensional infinite lattice is
always $d$-exact up to a constant form; $D$-forms are exceptional
because any $D$-form is $d$-closed. Moreover, the lemma asserts that
the locality is preserved between the original form and its
``ancestor.''
\proclaim Algebraic Poincar\'e lemma.
Let $\eta$~be a $p$-form on a $D$-dimensional infinite lattice that
depends smoothly and locally on the gauge potential~$A_\mu^a$ and on
the ghost field~$c^a$. Then
$$
   d\eta(n)=0\Leftrightarrow\eta(n)=d\chi(n)+{\cal L}(n)\,d^Dx+B,
\eqn\fivexone
$$
where $B$~is a constant $p$-form and the $(p-1)$-form $\chi$ and the
function~${\cal L}$ depend smoothly and locally on the gauge
potential and on the ghost field. The function~${\cal L}$ satisfies
$$
   \sum_n\delta{\cal L}(n)\neq0,
\eqn\fivextwo
$$
for a certain local variation~$\delta$ of the gauge potential and the
ghost field.

\noindent
{\it Note}. The term ${\cal L}\,d^Dx$ in~eq.~\fivexone\ represents a
non-topological part in the $D$-form~$\eta$. In other words, a
$D$-form~$\eta_{\rm top.}$ that is topological,
$\sum_n\delta\eta_{\rm top.}(n)=0$ for an arbitrary local variation,
is always $d$-exact up to a constant form.

\noindent
{\it Proof}. We define $\eta_t$ by rescaling fields as
$A_\mu^a\to tA_\mu^a$ and $c^a\to tc^a$. Then, since~$\eta$ depends
smoothly on~$A_\mu^a$ and on~$c^a$,
$$
\eqalign{
   \eta(n)&=\eta(n)_{t=0}
   +\int\nolimits_0^1dt\,{\partial\eta(n)_t\over\partial t}
\cr
   &=\eta(n)_{t=0}+\sum_{n'}
   \Bigl[A_\mu^a(n')\theta_\mu^a(n',n)+c^a(n')\kappa^a(n',n)\Bigr],
\cr
}
\eqn\fivexthree
$$
where
$$
   \theta_\mu^a(n',n)=\int\nolimits_0^1dt\,
   {\partial\eta(n)_t\over\partial tA_\mu^a(n')},\qquad
   \kappa^a(n',n)=\int\nolimits_0^1dt\,
   {\partial\eta(n)_t\over\partial tc^a(n')}.
\eqn\fivexfour
$$
Eq.~\fivexfour\ implies
$$
   d\theta_\mu^a(n',n)=d\kappa^a(n',n)=0,
\eqn\fivexfive
$$
because $d\eta=0$ for arbitrary configurations. Moreover,
since~$\eta$ depends locally on~$A_\mu^a$ and on~$c^a$,
$\theta_\mu^a(n',n)$ and $\kappa^a(n',n)$ decay exponentially as
$|n-n'|\to\infty$. This allows us to apply L\"uscher's Poincar\'e
lemma~[\LUS] for~$p<D$ to~eq.~\fivexfive\ which asserts that there
exist forms $\Theta_\mu^a$ and~$K^a$ such that
$$
   \theta_\mu^a(n',n)=d\Theta_\mu^a(n',n),\qquad
   \kappa^a(n',n)=dK^a(n',n).
\eqn\fivexsix
$$
These forms $\Theta_\mu^a(n',n)$ and~$K^a(n',n)$ also decay
exponentially as $|n-n'|\to\infty$~[\LUS]. Substituting this into
eq.~\fivexthree, we have $\eta=d\chi+B$, where $B=\eta_{t=0}$, and
$$
   \chi(n)=\sum_{n'}
   \Bigl[A_\mu^a(n')\Theta_\mu^a(n',n)+c^a(n')K^a(n',n)\Bigr].
\eqn\fivexseven
$$
{}From the locality property of $\Theta_\mu^a(n',n)$ and
of~$K^a(n',n)$~[\LUS], one can easily see~[\FUJ] that $\chi(n)$ is a
local field. Also the smoothness is preserved in the
construction~\fivexseven. In this way, the lemma~\fivexone\ is
established for~$p<D$.

For~$p=D$, $d\eta=0$ is a trivial statement and thus we
decompose~$\eta$ as
$$
   \eta=\eta_{\rm top.}+{\cal L}\,d^Dx,
\eqn\fivexeight
$$
where $\sum_n\delta\eta_{\rm top.}(n)=0$ for an arbitrary local
variation. Then $\theta_\mu^a$ and~$\kappa^a$
in~eq.~\fivexfour\ defined from~$\eta_{\rm top.}$ satisfy
$$
   \sum_n\theta_\mu^a(n',n)=\sum_n\kappa^a(n',n)=0.
\eqn\fivexnine
$$
Then L\"uscher's Poincar\'e lemma for~$p=D$~[\LUS] asserts that there
exist $\Theta_\mu^a$ and~$K^a$ which satisfy eq.~\fivexsix. The rest
is the same as for~$p<D$ and we have $\eta_{\rm top.}=d\chi+B$.\QED

\section{Abelian BRS cohomology}
\proclaim Abelian BRS cohomology.
Let $X$~be a form on infinite hypercubic lattice that depends smoothly
and locally on the gauge potential and on the ghost field. Then,
$$
   sX(n)=0\Leftrightarrow X(n)
   =C^{a_1}(n)\cdots C^{a_g}(n)X_0^{[a_1\cdots a_g]}[\{F_i\};n]+sY(n),
\eqn\fivexten
$$
where the form $X_0^{[a_1\cdots a_g]}(n)$ depends smoothly and locally
only on the abelian field strength~$F_{\mu\nu}^a$. The form~$Y(n)$
depends smoothly and locally on the gauge potential and on the ghost
field. In particular, differences of the ghost field can appear only
in the BRS trivial part~$sY$.

\noindent
{\it Note}. The form~$X_0^{[a_1\cdots a_g]}$ is totally antisymmetric
on the upper indices because ghost 1-forms $C^a$ simply anticommute
with each other. $X_0^{[a_1\cdots a_g]}(n)$ depends only on the field
strength~$F_{\mu\nu}^a(n)$ and its differences, such as
$\Delta_\mu F_{\nu\rho}^a(n)$, $\Delta_\mu^*F_{\nu\rho}^a(n)$,
$\Delta_\mu\Delta_\nu F_{\rho\sigma}^a(n)$ and so on; obviously
$X_0^{[a_1\cdots a_g]}$ is gauge invariant. In what follows, we denote
as~$X_0^{[a_1\cdots a_g]}[\{F_i\}]$ to indicate this particular
dependence on the field strength, including smoothness and locality
of the dependence.

\noindent
{\it Proof}. The proof of the abelian BRS cohomology for a
single~$U(1)$ case~[\FUJ] can be repeated by simply supplementing
the gauge potential~$A_\mu$ and the ghost field~$c$ by another
index~$a$. Thus we do not reproduce it here to save the space.\QED

\section{Covariant Poincar\'e lemma}
As in the continuum theory~[\BRA], the following covariant
Poincar\'e lemma is crucial to determine general nontrivial local
solutions to the consistency condition. This lemma for a single~$U(1)$
case $G=U(1)$ was given in~ref.~[\FUJ]. It turns out that, however,
its extension to multi-$U(1)$ cases is not trivial, due to the reason
which will be explained after the proof. In fact, we have at present
only the following cumbersome proof that works only for 4- or lower
dimensional lattice.

\proclaim Covariant Poincar\'e lemma.
On a 4-dimensional infinite hypercubic lattice, if the
$p$-form~$\alpha_p[\{F_i\}]$ is $d$-closed for $p<4$, or if
$\alpha_4[\{F_i\}]=d\chi_3+B_4$ where $B_4$~is a constant 4-form, then
$\alpha_p$~is of the structure
$$
   \alpha_p[\{F_i\};n]
   =d\alpha_{p-1}[\{F_i\};n]+B_p+F^a(n)B_{p-2}^a
   +F^a(n)F^b(n)B_{p-4}^{(ab)},
\eqn\fivexeleven
$$
where $F^a$ is the field strength 2-form and $B$'s are constant forms.

\noindent
{\it Note}. Here all expressions are written in terms of the
noncommutative differential calculus.

\noindent
{\it Proof}. We prove the lemma step by step from 0-form~$p=0$ until
4-form~$p=4$.

\noindent
\undertext{For $p=0$}. The lemma trivially holds by the algebraic
Poincar\'e lemma~\fivexone. Namely, the $d$-closed 0-form~$\alpha_0$
must be a constant $\alpha_0=B_0$.

\noindent
\undertext{For $p=1$}. By the algebraic Poincar\'e lemma, the
$d$-closed 1-form~$\alpha_1$ is $d$-exact up to a constant 1-form.
Also $\alpha_1$ is $s$-closed because it is a function of the field
strength. Namely,
$$
   \alpha_1=d\chi_0^0+B_1,\qquad s\alpha_1=0.
\eqn\fivextwelve
$$
Since these equations imply $s\alpha_1=sd\chi_0^0=-ds\chi_0^0=0$, the
algebraic Poincar\'e lemma asserts that
$$
   s\chi_0^0=0,
\eqn\fivexthirteen
$$
where we have used the fact that the right hand side cannot be a
constant. The solution to this equation is given by the abelian BRS
cohomology~\fivexten\ for the $g=0$~case:
$$
   \chi_0^0=\omega_0[\{F_i\}],
\eqn\fivexfourteen
$$
and thus eq.~\fivextwelve\ shows that the lemma holds for $p=1$:
$$
   \alpha_1=d\omega_0[\{F_i\}]+B_1.
\eqn\fivexfifteen
$$

\noindent
\undertext{For $p=2$}. In this case, from the algebraic Poincar\'e
lemma, we have
$$
   \alpha_2=d\chi_1^0+B_1,\qquad s\alpha_2=0,
\eqn\fivexsixteen
$$
and, in a similar way as the $p=1$ case, these lead to the following
descent equations
$$
   s\chi_1^0=d\chi_0^1,\qquad s\chi_0^1=0.
\eqn\fivexseventeen
$$
The general solution to the last equation is given by the abelian BRS
cohomology
$$
   \chi_0^1=C^a\omega_0^a[\{F_i\}]+s\beta_0.
\eqn\fivexeighteen
$$
We may, however, absorb $\beta_0$ in redefinition of~$\chi_0^1$
and~$\chi_1^0$,
$$
   \chi_0^1\to\chi_0^1+s\beta_0,\qquad\chi_1^0\to\chi_1^0-d\beta_0,
\eqn\fivexnineteen
$$
without changing $\alpha_2$. We can therefore take
$\chi_0^1=C^a\omega_0^a$. Then the first equation
in~eq.~\fivexseventeen\ reads
$$
\eqalign{
   s\chi_1^0&=dC^a\omega_0^a-C^ad\omega_0^a
\cr
   &=-s(A^a\omega_0^a)-C^ad\omega_0^a.
\cr
}
\eqn\fivextwenty
$$
Now consider a special configuration of the ghost field
$c^a(n)\to c^a={\rm const}$. Then the consistency
of~eq.~\fivextwenty\ requires
$$
   d\omega_0^a=0,\qquad s(\chi_1^0+A^a\omega_0^a)=0,
\eqn\fivextwentyone
$$
because $s{\rm(something)}$ is proportional to differences of the
ghost fields such as $c^a(n+\widehat\mu)-c^a(n)$. Note that
$\omega_0^a$ does not depend on the ghost field. The solution to the
first equation of~eq.~\fivextwentyone\ is given by the present lemma
for $p=0$, which we have shown above:
$$
   \omega_0^a=B_0^a\quad{\rm (const.)},
\eqn\fivextwentytwo
$$
and then the second relation of~eq.~\fivextwentyone\ implies
$$
   \chi_1^0=-A^aB_0^a+\omega_1[\{F_i\}],
\eqn\fivextwentythree
$$
by the BRS cohomology. Going back to the original
relation~\fivexsixteen, we have
$$
   \alpha_2=-F^aB_0^a+d\omega_1[\{F_i\}]+B_2,
\eqn\fivextwentyfour
$$
because $dA^a=F^a$. This proves the lemma for~$p=2$.

\noindent
\undertext{For $p=3$}. In this case, the counterparts
of~eqs.~\fivexsixteen\ and~\fivexseventeen\ are
$$
   \alpha_3=d\chi_2^0+B_3,\qquad s\alpha_3=0,
\eqn\fivextwentyfive
$$
and
$$
   s\chi_2^0=d\chi_1^1,\qquad s\chi_1^1=d\chi_0^2,\qquad
   s\chi_0^2=0.
\eqn\fivextwentysix
$$
The solution to the last equation is (we have absorbed the BRS trivial
part as eq.~\fivexnineteen)
$$
   \chi_0^2=C^aC^b\omega_0^{[ab]}[\{F_i\}],
\eqn\fivextwentyseven
$$
where $\omega_0^{[ab]}$ is antisymmetric under~$a\leftrightarrow b$.
At this stage, it is quite convenient to introduce the symmetrization
symbol defined by
$$
   \sym(X_1X_2\cdots X_N)=\sum_\sigma{1\over N!}\,\epsilon_\sigma
   X_{\sigma(1)}X_{\sigma(2)}\cdots X_{\sigma(N)},
\eqn\fivextwentyeight
$$
where the summation is taken over all permutations~$\sigma$. The sign
factor~$\epsilon_\sigma$ is defined as the signature arising when the
product $X_1\cdots X_N$ is converted to the order
$X_{\sigma(1)}X_{\sigma(2)}\cdots X_{\sigma(N)}$ by regarding all
$X_i$'s as {\it ordinary\/} forms (i.e., the form basis $dx_\mu$
simply commutes with functions). Using the symmetrization symbol,
eq.~\fivextwentyseven\ is trivially written as
$$
   \chi_0^2=\sym(C^aC^b)\omega_0^{[ab]},
\eqn\fivextwentynine
$$
and then the second relation of~eq.~\fivextwentysix\ reads
$$
\eqalign{
   s\chi_1^1&=2\sym(dC^aC^b)\omega_0^{[ab]}
   +\sym(C^aC^b)d\omega_0^{[ab]}
\cr
   &=s\Bigl[-2\sym(A^aC^b)\omega_0^{[ab]}\Bigr]
   +\sym(C^aC^b)d\omega_0^{[ab]}.
\cr
}
\eqn\fivexthirty
$$
Let us now consider the special configuration $c^a(n)\to{\rm const}$.
As for eq.~\fivextwenty, the consistency of~eq.~\fivexthirty\ requires
$$
   d\omega_0^{[ab]}=0,\qquad
   s\Bigl[\chi_1^1+2\sym(A^aC^b)\omega_0^{[ab]}\Bigr]=0.
\eqn\fivexthirtyone
$$
The general solution to the first equation is $\omega_0^{[ab]}=%
B_0^{[ab]}$ and then the second equation implies (by the BRS
cohomology) $\chi_1^1=-2\sym(A^aC^b)B_0^{[ab]}+%
C^a\omega_1^a[\{F_i\}]$. Substituting these into the first relation
of eq.~\fivextwentysix, we have
$$
   s\chi_2^0=s\Bigl[\sym(A^aA^b)B_0^{[ab]}-A^a\omega_1^a\Bigr]
   -2\sym(F^aC^b)B_0^{[ab]}-C^ad\omega_1^a.
\eqn\fivexthirtytwo
$$
We again consider the configuration $c^a(n)\to{\rm const}$. Then
eq.~\fivexthirtytwo\ requires
$$
   2F^aB_0^{[ab]}+d\omega_1^b=0,\qquad
   s\Bigl[\chi_2^0-\sym(A^aA^b)B_0^{[ab]}+A^a\omega_1^a\Bigr]=0.
\eqn\fivexthirtythree
$$
In deriving the first relation, we have noted the fact that the
{\it constant\/} ghost form $C^b$ and the 2-form $F^a$ simply commute,
and thus $C^b$ can be factored out from the equation. Next we consider
a configuration $F_{\mu\nu}^a(n)\to{\rm const}$. Since $\omega_1^a$
depends only on the field strength, we see that the
constant~$B_0^{[ab]}$ must vanish for the consistency
of~eq.~\fivexthirtythree\foot{Later, we apply the covariant Poincar\'e
lemma to the case~II above, by regarding components of the nonabelian
gauge potential $A_\mu(n)=\sum_a A_\mu^a(n)T^a$ as if they were the
abelian gauge potential. In this case, it is impossible to
take~$F_{\mu\nu}^a(n)={\rm const.}$ while keeping the range
of~$A_\mu^a(n)$ as~eq.~\twoxnine. However, it is possible to take
$F_{\mu\nu}^a(n)={\rm const.}=O(1/R)$ inside of a block of size~$R$.
The term~$d\omega_1^b$ then behaves as~$\sim\exp(-\alpha R)$ because
the dependence of~$\omega_1^b$ is local. Since the first relation
of~eq.~\fivexthirtythree\ holds for arbitrary~$R$, this implies that
each term has to vanish separately.} and thus
$$
   d\omega_1^a=0\Rightarrow
   \omega_1^a=d\omega_0^a[\{F_i\}]+B_1^a,
\eqn\fivexthirtyfour
$$
by the present lemma for~$p=1$. Substituting this into the second
relation of~eq.~\fivexthirtythree\ and using the first
equation~\fivextwentyfive, we have
$$
\eqalign{
   \alpha_3&=-F^ad\omega_0^a-F^aB_1^a+d\omega_2+B_3
\cr
   &=d(-F^a\omega_0^a+\omega_2)-F^aB_1^a+B_3,
\cr
}
\eqn\fivexthirtyfive
$$
where we have used the Bianchi identity~$dF^a=0$. This shows the lemma
for~$p=3$.

\noindent
\undertext{For $p=4$}. Similarly as the above cases, we have
$$
   \alpha_4=d\chi_3^0+B_4,\qquad s\alpha_4=0,
\eqn\fivexthirtysix
$$
and
$$
   s\chi_3^0=d\chi_2^1,\qquad s\chi_2^1=d\chi_1^2,\qquad
   s\chi_1^2=d\chi_0^3,\qquad s\chi_0^3=0.
\eqn\fivexthirtyseven
$$
The solution to the last equation is given by $\chi_0^3=%
\sym(C^aC^bC^c)\omega_0^{[abc]}[\{F_i\}]$ and then the third relation
of~eq.~\fivexthirtyseven\ reads
$$
   s\chi_1^2=
   s\Bigl[-3\sym(A^aC^bC^c)\omega_0^{[abc]}\Bigr]
   -\sym(C^aC^bC^c)d\omega_0^{[abc]}.
\eqn\fivexthirtyeight
$$
The consistency for $c^a(n)\to{\rm const.}$ requires
$\omega_0^{[abc]}=B_0^{[abc]}$ (const.) and thus
$$
   \chi_1^2=-3\sym(A^aC^bC^c)B_0^{[abc]}
   +\sym(C^aC^b)\omega_1^{[ab]}[\{F_i\}],
\eqn\fivexthirtynine
$$
and the second equation of~eq.~\fivexthirtyseven\ becomes
$$
\eqalign{
   s\chi_2^1&=s\Bigl[-3\sym(A^aA^bC^c)B_0^{[abc]}
   -2\sym(A^aC^b)\omega_1^{[ab]}\Bigr]
\cr
   &\qquad-3\sym(F^aC^bC^c)B_0^{[abc]}+\sym(C^aC^b)d\omega_1^{[ab]}.
\cr
}
\eqn\fivexforty
$$
Setting $c^a(n)\to{\rm const.}$ in this equation and then setting
$F_{\mu\nu}^a(n)\to{\rm const.}$, we see that $B_0^{[abc]}=0$ and
$d\omega_1^{[ab]}=0$. The present lemma for~$p=1$ then asserts that
$\omega_1^{[ab]}=d\omega_0^{[ab]}[\{F_i\}]+B_1^{[ab]}$. The general
structure of $\chi_2^1$ is therefore given by
$$
   \chi_2^1=-2\sym(A^aC^b)(d\omega_0^{[ab]}+B_1^{[ab]})
   +C^a\omega_2^a[\{F_i\}].
\eqn\fivexfortyone
$$
Substituting this into the first relation of eq.~\fivexthirtyseven, we
have
$$
\eqalign{
   s\chi_3^0
   &=s\Bigl[\sym(A^aA^b)(d\omega_0^{[ab]}+B_1^{[ab]})
   -A^a\omega_2^a\Bigr]
\cr
   &\qquad-2\sym(F^aC^b)(d\omega_0^{[ab]}+B_1^{[ab]})-C^ad\omega_2^a.
\cr
}
\eqn\fivexfortytwo
$$
The consistency for $c^a(n)\to{\rm const.}$ requires
$$
   2F^a(d\omega_0^{[ab]}+B_1^{[ab]})+d\omega_2^b=0,
\eqn\fivexfortythree
$$
and the consistency for $F_{\mu\nu}^a(n)\to{\rm const.}$,
$$
   B_1^{[ab]}=0,\qquad 2F^ad\omega_0^{[ab]}+d\omega_2^b=0.
\eqn\fivexfortyfour
$$
The last equation can be written as
$d(\omega_2^a-2F^b\omega_0^{[ab]})=0$, and then the present lemma
for~$p=2$ asserts that
$$
   \omega_2^a=2F^b\omega_0^{[ab]}+d\omega_1^a[\{F_i\}]
   +B_2^a+F^bB_0^{ab}.
\eqn\fivexfortyfive
$$
Note that $B_0^{ab}$ is not necessarily symmetric
under~$a\leftrightarrow b$ at this stage. From this it is not
difficult to see that eq.~\fivexfortytwo\ yields
$$
\eqalign{
   s\chi_3^0
   &=sd\Bigl[\sym(A^aA^b)\omega_0^{[ab]}\Bigr]
   +s\Bigl[-A^a(d\omega_1^a+B_2^a+F^bB_0^{ab})\Bigr]
\cr
   &\qquad+d\Bigl\{
   2\Bigl[\sym(F^aC^b)+C^aF^b\Bigr]\omega_0^{[ab]}\Bigr\}.
\cr
}
\eqn\fivexfortysix
$$

We have now arrived at the final stage which requires special
consideration. In eq.~\fivexfortysix, the last term on the right hand
side is not manifestly $s$-exact. So define
$$
   \varphi_2^1=2\Bigl[\sym(F^aC^b)+C^aF^b\Bigr]\omega_0^{[ab]}
   =(C^aF^b-F^bC^a)\omega_0^{[ab]}.
\eqn\fivexfortyseven
$$
In the context of ordinary differential calculus, $\varphi_2^1$
identically vanishes because $C^a$ and~$F^b$ commute with each other.
However we cannot simply throw away $\varphi_2^1$ in the context of
noncommutative differential calculus. We first note~$s\varphi_2^1=0$.
Also, when $c^a(n)\to{\rm const.}$, $C^a$ and~$F^b$ commute and
$\varphi_2^1=0$ as noted above. Therefore $\varphi_2^1\propto%
\Delta_\mu c^a$. These facts combined with the BRS
cohomology~\fivexten\ show that $\varphi_2^1$ is $s$-trivial,
$\varphi_2^1=sY_2$ (actually, otherwise eq.~\fivexfortysix\ becomes
inconsistent). In fact, by noting the noncommutative rule~\fourxfour,
one finds
$$
   \varphi_2^1=-\bigl[\Delta_\mu c^a(n)+\Delta_\nu c^a(n)
   +\Delta_\mu\Delta_\nu c^a(n)\bigr]d\theta
   \,{1\over2}F_{\mu\nu}^b(n)\,dx_\mu dx_\nu\,\omega_0^{[ab]}
   =sY_2,
\eqn\fivexfortyeight
$$
where
$$
   Y_2=-\bigl[A_\mu^a(n)+A_\nu^a(n)+\Delta_\nu A_\mu^a(n)\bigr]
   \,{1\over2}F_{\mu\nu}^b(n)\,dx_\mu dx_\nu\,
   \omega_0^{[ab]}.
\eqn\fivexfortynine
$$
Therefore eq.~\fivexfortysix\ gives
$$
   \chi_3^0=-A^a(d\omega_1^a+B_2^a+F^bB_0^{ab})
   +\omega_3[\{F_i\}]
   +d\Bigl[\sym(A^aA^b)\omega_0^{[ab]}-Y_2\Bigr],
\eqn\fivexfifty
$$
and from the first equation~\fivexthirtysix, we have
$$
   \alpha_4=d(-F^a\omega^a_1+\omega_3)-F^aB_2^a-F^aF^bB_0^{ab}+B_4.
\eqn\fivexfiftyone
$$

Finally, we show that the term which is proportional to the
antisymmetric part of~$B_0^{ab}$ under $a\leftrightarrow b$, and which
again vanishes in ordinary differential calculus, can be expressed
as~$d\varphi_3(\{F_i\})$. First note that $F^aF^bB_0^{[ab]}=%
d\varphi_3$ where
$$
   \varphi_3={1\over2}(A^aF^b-F^bA^a)B_0^{[ab]}-{1\over2}dY_2,
\eqn\
$$
and $Y_2$ in the second term is defined by~$\omega_0^{[ab]}\to%
B_0^{[ab]}$ in~eq.~\fivexfortynine. Of course, the last
term~$-dY_2/2$ does dot contribute to~$F^aF^bB_0^{[ab]}$, but it makes
$\varphi_3$ gauge invariant. In fact,
$$
   s\varphi_3=-{1\over2}(dC^aF^b-F^bdC^a)B_0^{[ab]}+{1\over2}dsY_2
   =0,
\eqn\fivexfiftythree
$$
where use of eqs.~\fivexfortyeight\ and~\fivexfortyseven\ has been
made. More explicitly, after some calculation with use of the Bianchi
identity, we have
$$
\eqalign{
   \varphi_3(n)&={1\over8}\bigl[
   F_{\alpha\beta}^a(n)F_{\beta\gamma}^b(n)
   +F_{\alpha\beta}^a(n+\widehat\gamma)F_{\beta\gamma}^b(n)
\cr
   &\qquad\quad
   +F_{\alpha\beta}^a(n)F_{\beta\gamma}^b(n+\widehat\alpha)
   +F_{\alpha\beta}^a(n+\widehat\gamma)
   F_{\beta\gamma}^b(n+\widehat\alpha)\bigr]
   dx_\alpha\,dx_\beta\,dx_\gamma B_0^{[ab]}.
\cr
}
\eqn\fivexfiftyfour
$$
This establishes the lemma for~$p=4$.\QED

If one repeats the above argument for~$p=5$ (assuming that the
dimension of the lattice is greater than 4), the treatment becomes
much more involved due to the noncommutativity of forms. Because of
this, we could not find an iterative formula for $\alpha_p$ with
general~$p$, unlike the treatment in the continuum theory~[\BRA]. This
fact suggests that our noncommutative differential calculus is not
powerful enough and there must exist another hidden algebraic
structure. This is an interesting problem although we do not
investigate it here. Of course, the proof in this subsection is
sufficient for applications on 4-dimensional lattices.

\section{Topological fields in the abelian theory}
Once the above three lemmas are established, it is straightforward to
show the following theorem which generalizes the theorem
of~ref.~[\LUS] to multi-$U(1)$ cases.

\proclaim Theorem.
Let $q(n)$~be a gauge invariant field on a 4-dimensional infinite
hypercubic lattice that depends smoothly and locally on the abelian
gauge potential~$A_\mu^a$. Suppose that~$q(n)$ is topological, namely
$$
   \sum_n\delta q(n)=0,
\eqn\fivexfiftyfive
$$
for an arbitrary local variation of the gauge potential. Then $q(n)$
is of the form
$$
   q(n)=\alpha+\beta_{\mu\nu}^aF_{\mu\nu}^a(n)
   +\gamma^{(ab)}\varepsilon_{\mu\nu\rho\sigma}F_{\mu\nu}^a(n)
   F_{\rho\sigma}^b(n+\widehat\mu+\widehat\nu)+\Delta_\mu^*k_\mu(n),
\eqn\fivexfiftysix
$$
where the current~$k_\mu(n)$ depends smoothly and locally only on the
field strength and thus is gauge invariant.

\noindent
{\it Proof}. We multiply the volume form $d^4x$ to $q(n)$ and define
the 4-form $Q_4=q\,d^4x$. $Q_4$~is gauge invariant $sQ_4=0$ and thus,
{}from the BRS cohomology~\fivexten, $Q_4=Q_4(\{F_i\})$. From the
algebraic Poincar\'e lemma~\fivexone, on the other hand, $Q_4=%
d\chi_3+B_4$ because the 4-form~$Q_4$ is
topological~$\sum_n\delta Q_4(n)=0$ from the
assumption~\fivexfiftyfive. From these, we can apply the covariant
Poincar\'e lemma~\fivexeleven\ to $Q_4$ which yields
$$
   q(n)d^4x=Q_4(n)=B_4+F^a(n)B_2^a+F^a(n)F^b(n)B_0^{(ab)}
   +d\alpha_3(n).
\eqn\fivexfiftyseven
$$
Finally, we factor out the volume form~$d^4x$ from the both sides of
this equation. Noting the noncommutative rule~\fourxfour, we have
eq.~\fivexfiftysix.\QED

\noindent
{\it Note}. The third term of~eq.~\fivexfiftysix\ is a total
difference on the lattice and thus in fact satisfies the topological
property~\fivexfiftyfive:
$$
   \varepsilon_{\mu\nu\rho\sigma}F_{\mu\nu}^a(n)
   F_{\rho\sigma}^b(n+\widehat\mu+\widehat\nu)
   =4\varepsilon_{\mu\nu\rho\sigma}\Delta_\mu\bigl[
   A_\nu^a(n)\Delta_\rho A_\sigma^b(n+\widehat\nu)\bigr].
\eqn\fivexfiftyeight
$$
This relation can easily be derived from the relation~$F^aF^b=%
d(A^aF^b)$ which is valid in the context of noncommutative
differential calculus. See ref.~[\FUJ].

\chapter{Nontrivial anomalies in the abelian theory}
Because of the nilpotency~$\delta_B^2=0$, any functional of the form
${\cal A}=\delta_B{\cal B}$ is a solution to~eq.~\onexthree. If the
functional~${\cal B}$ is {\it local}, such an anomaly can be removed
by the redefinition of the effective action~$\ln\Det M'\to%
\ln\Det M'-{\cal B}$ which does not change the physical content of
the theory. Therefore the solution to the consistency
condition~\onexthree\ of the form~${\cal A}=\delta_B{\cal B}$ with a
{\it local\/} functional~${\cal B}$ will be referred as trivial or BRS
trivial.

\section{Nontrivial local solutions}
In this subsection, we study the structure of local solutions to the
consistency condition~\onexthree\ in the abelian theory~$G=U(1)^N$.
The BRS transformation is given by~eq.~\fourxseven. The ghost number
of the solution is not restricted. We will find a very close analogue
to the solutions in the continuum theory~[\BRA]. In the terminology
of~ref.~[\DUB], our result gives ${\rm H}(\delta|d)$ in abelian
theories.

We seek the solution~${\cal A}$ by regarding ${\cal A}$ as a smooth
and local functional of the gauge potential~$A_\mu^a$ and the ghost
field~$c^a$. Since eq.~\onexthree\ must hold for arbitrary
configurations of~$A_\mu^a$ and~$c^a$, we have the variational
equation
$$
   \delta\delta_B{\cal A}
   =\sum_n\biggl\{
   \delta A_\mu^a(n)\delta_B{\partial{\cal A}\over\partial A_\mu^a(n)}
   -\delta c^a(n)\biggl[
   \delta_B{\partial{\cal A}\over\partial c^a(n)}
   +\Delta_\mu^*{\partial{\cal A}\over\partial A_\mu^a(n)}\biggr]
   \biggr\}=0,
\eqn\sixxone
$$
where $\delta\delta_BA_\mu^a=\Delta_\mu\delta c^a$
and~$\delta\delta_B c^a=0$ have been used. The coefficients of the
variations $\delta A_\mu^a(n)$ and~$\delta c^a(n)$ have to vanish
separately:
$$
   \delta_B{\partial{\cal A}\over\partial A_\mu^a(n)}=0,\qquad
   \delta_B{\partial{\cal A}\over\partial c^a(n)}
   +\Delta_\mu^*{\partial{\cal A}\over\partial A_\mu^a(n)}=0.
\eqn\sixxtwo
$$
Since ${\cal A}$ is local, $\partial{\cal A}/\partial A_\mu^a(n)$ is a
local field. Then the abelian BRS cohomology~\fivexten\ gives the
general solution to the first equation with the ghost number~$g$,
$$
   {\partial{\cal A}\over\partial A_\mu^a(n)}
   =c^{a_1}(n)\cdots c^{a_g}(n)
   \omega_\mu^{a[a_1\cdots a_g]}(n)+\delta_BY_\mu^a(n),
\eqn\sixxthree
$$
where $\omega_\mu^{a[a_1\cdots a_g]}$ depends only on the field
strength. The second relation of~eq.~\sixxtwo\ then reads,
$$
\eqalign{
   &\delta_B
   \biggl[{\partial{\cal A}\over\partial c^a(n)}
   +\Delta_\mu^*Y_\mu^a(n)\biggr]
   =-\Delta_\mu^*\Bigl[
   c^{a_1}(n)\cdots c^{a_g}(n)\omega_\mu^{a[a_1\cdots a_g]}(n)\Bigr]
\cr
   &=-c^{a_1}(n)\cdots c^{a_g}(n)\omega_\mu^{a[a_1\cdots a_g]}(n)
   +c^{a_1}(n-\widehat\mu)\cdots c^{a_g}(n-\widehat\mu)
   \omega_\mu^{a[a_1\cdots a_g]}(n-\widehat\mu)
\cr
   &=\delta_B\Biggl[-\sum_{i=1}^g{g\choose i}
   A_\mu^{a_1}(n-\widehat\mu)\delta_BA_\mu^{a_2}(n-\widehat\mu)\cdots
   \delta_BA_\mu^{a_i}(n-\widehat\mu)
\cr
   &\qquad\qquad\qquad\qquad\qquad\qquad\qquad\qquad\qquad
   \times c^{a_{i+1}}(n-\widehat\mu)\cdots c^{a_g}(n-\widehat\mu)
   \omega_\mu^{a[a_1\cdots a_g]}(n)\Biggr]
\cr
   &\qquad-c^{a_1}(n-\widehat\mu)\cdots c^{a_g}(n-\widehat\mu)
   \Delta_\mu^*\omega_\mu^{a[a_1\cdots a_g]}(n),
\cr
}
\eqn\sixxfour
$$
where we have used $c^a(n)=c^a(n-\widehat\mu)+%
\delta_BA_\mu^a(n-\widehat\mu)$ to pass from the second line to the
third line. Considering the consistency of the above equation under
$c^a(n)\to{\rm const.}$, we have
$$
   \Delta_\mu^*\omega_\mu^{a[a_1\cdots a_g]}(n)=0,
\eqn\sixxfive
$$
and then again from the BRS cohomology,
$$
\eqalign{
   {\partial{\cal A}\over\partial c^a(n)}
   &=-\Delta_\mu^*Y_\mu^a(n)-\sum_{i=1}^g{g\choose i}
   A_\mu^{a_1}(n-\widehat\mu)\delta_BA_\mu^{a_2}(n-\widehat\mu)\cdots
   \delta_BA_\mu^{a_i}(n-\widehat\mu)
\cr
   &\qquad\qquad\qquad\qquad\qquad\qquad\qquad
   \times c^{a_{i+1}}(n-\widehat\mu)\cdots c^{a_g}(n-\widehat\mu)
   \omega_\mu^{a[a_1\cdots a_g]}(n)
\cr
   &\qquad+c^{a_1}(n)\cdots c^{a_{g-1}}(n)X^{a[a_1\cdots a_{g-1}]}(n)
   +\delta_BY^a(n),
\cr
}
\eqn\sixxsix
$$
where $X^{a[a_1\cdots a_{g-1}]}$ depends only on the field strength.

The functional~${\cal A}$ can be reconstructed from its
variations~\sixxthree\ and~\sixxsix\ as follows. We
introduce~${\cal A}_t$ by rescaling variables as $A_\mu^a\to%
tA_\mu^a$ and~$c^a\to tc^a$. Noting ${\cal A}_{t=0}=0$
for~$g>0$,\foot{For~$g=0$, the following expressions hold by simply
adding a constant~${\cal A}_{t=0}$.} we have
$$
   {\cal A}=\int\nolimits_0^1dt\,{\partial{\cal A}_t\over\partial t}
   =\int\nolimits_0^1dt\,\sum_n\biggl[
   A_\mu^a(n){\partial{\cal A}_t\over\partial tA_\mu^a(n)}
   +c^a(n){\partial{\cal A}_t\over\partial tc^a(n)}\biggr].
\eqn\sixxseven
$$
After substituting eqs.~\sixxthree\ and~\sixxsix\ and shifting the
coordinate~$s\to s+\widehat\mu$, this yields
$$
\eqalign{
   {\cal A}&=\sum_n\biggl\{\biggl[
   A_\mu^{a_0}(n)c^{a_1}(n)\cdots c^{a_g}(n)
\cr
   &\qquad\qquad
   -c^{a_0}(n+\widehat\mu)\sum_{i=1}^g{g\choose i}
   A_\mu^{a_1}(n)\delta_BA_\mu^{a_2}(n)\cdots\delta_BA_\mu^{a_i}(n)
   c^{a_{i+1}}(n)\cdots c^{a_g}(n)\biggr]
\cr
   &\qquad\qquad\qquad\qquad\qquad\qquad\qquad\qquad\qquad\qquad
   \qquad\qquad\qquad
   \times\widetilde\omega_\mu^{a_0[a_1\cdots a_g]}(n)
\cr
   &\qquad\qquad+c^{a_1}(n)\cdots c^{a_g}(n)
   \widetilde X^{[a_1\cdots a_g]}(n)\biggr\}
\cr
   &\qquad+\delta_B\sum_n\Bigl[
   A_\mu^a(n)\widetilde Y_\mu^a(n)-c^a(n)\widetilde Y^a(n)\Bigr],
\cr
}
\eqn\sixxeight
$$
where the following abbreviations have been introduced
$$
\eqalign{
   &\widetilde\omega_\mu^{a[a_1\cdots a_g]}
   =\int\nolimits_0^1dt\,t^g\omega_\mu^{a[a_1\cdots a_g]}{}_t,\qquad
   \widetilde X^{[a_1\cdots a_g]}
   =\int\nolimits_0^1dt\,t^{g-1}X^{[a_1\cdots a_g]}{}_t,
\cr
   &\widetilde Y_\mu^a=\int\nolimits_0^1dt\,Y_\mu^a{}_t,\qquad
   \widetilde Y^a=\int\nolimits_0^1dt\,Y^a{}_t.
\cr
}
\eqn\sixxnine
$$
Note that $\Delta_\mu^*\widetilde\omega_\mu^{a[a_1\cdots a_g]}=0$
{}from~eq.~\sixxfive\ and that all these fields are local from the
above construction. In particular,
$\widetilde\omega_\mu^{a[a_1\cdots a_g]}$
and~$\widetilde X^{[a_1\cdots a_g]}$ depend only on the field
strength.

Eq.~\sixxeight\ provides the most general local solutions to the
consistency condition. Yet it contains trivial solutions in various
ways. First, by noting~$c^{a_0}(n+\widehat\mu)=%
c^{a_0}(n)+\delta_BA_\mu^{a_0}(n)$ and~$\delta_BA_\mu^{a_2}(n)=%
c^{a_2}(n+\widehat\mu)-c^{a_2}(n)$, it is easy to see that the
symmetric part of~$\widetilde\omega_\mu^{a_0[a_1\cdots a_g]}$
on~$a_0\leftrightarrow a_1$ contributes only to a BRS trivial part:
$$
\eqalign{
   &\Bigl[A_\mu^{a_0}(n)c^{a_1}(n)\cdots c^{a_g}(n)
   -gc^{a_0}(n+\widehat\mu)
   A_\mu^{a_1}(n)c^{a_2}(n)\cdots c^{a_g}(n)\Bigr]
   \widetilde\omega_\mu^{(a_0a_1)a_2\cdots a_g}(n)
\cr
   &=\delta_B\biggl[
   -{g\over2}A_\mu^{a_0}(n)A_\mu^{a_1}(n)c^{a_2}(n)\cdots c^{a_g}(n)
   \widetilde\omega_\mu^{(a_0a_1)a_2\cdots a_g}(n)\biggr],
\cr
}
\eqn\sixxten
$$
and
$$
\eqalign{
   &c^{a_0}(n+\widehat\mu)A_\mu^{a_1}(n)\delta_BA_\mu^{a_2}(n)
   \cdots\delta_BA_\mu^{a_i}(n)c^{a_{i+1}}(n)\cdots c^{a_g}(n)
   \widetilde\omega_\mu^{(a_0a_1)a_2\cdots a_g}(n)
\cr
   &=\delta_B\biggl[-{1\over2}
   A_\mu^{a_0}(n)A_\mu^{a_1}(n)c^{a_2}(n)\delta_BA_\mu^{a_3}(n)
   \cdots\delta_BA_\mu^{a_i}(n)c^{a_{i+1}}(n)\cdots c^{a_g}(n)
   \widetilde\omega_\mu^{(a_0a_1)a_2\cdots a_g}(n)\biggr].
\cr
}
\eqn\sixxeleven
$$
Therefore, for nontrivial solutions, we can assume that
$\widetilde\omega_\mu^{a_0[a_1\cdots a_g]}$~is antisymmetric
under the exchange $a_0\leftrightarrow a_1$, namely,
$\widetilde\omega_\mu^{a_0[a_1\cdots a_g]}$~is totally
antisymmetric $\widetilde\omega_\mu^{a_0\cdots a_g}=%
\widetilde\omega_\mu^{[a_0\cdots a_g]}$ in nontrivial solutions.

Henceforth we use the symbol~$\simeq$ to indicate the equivalence
relation modulo BRS trivial parts. The last term
of~eq.~\sixxeight\ is BRS trivial. Also, as noted above,
$\widetilde\omega_\mu^{a_0\cdots a_g}$ is totally antisymmetric in
nontrivial solutions. Then, by inserting $\delta_BA_\mu^{a_j}(n)=%
c^{a_j}(n+\widehat\mu)-c^{a_j}(n)$ into~eq.~\sixxeight, and after some
rearrangements, we have the following relatively simple expression
$$
\eqalign{
   {\cal A}&\simeq
   \sum_n\Biggl[
   \sum_{k=0}^gc^{a_1}(n)\cdots c^{a_k}(n)A_\mu^{a_0}(n)
   c^{a_{k+1}}(n+\widehat\mu)\cdots c^{a_g}(n+\widehat\mu)
   \widetilde\omega_\mu^{[a_0\cdots a_g]}(n)
\cr
   &\qquad\qquad
   +c^{a_1}(n)\cdots c^{a_g}(n)\widetilde X^{[a_1\cdots a_g]}(n)
   \Biggr].
\cr
}
\eqn\sixxtwelve
$$
This expression takes a particularly simple form in terms of the
noncommutative differential calculus. We introduce the dual 3-form
of~$\widetilde\omega_\mu$ by
$$
   \widetilde\omega_\mu^{[a_0\cdots a_g]}(n)
   ={1\over3!}\varepsilon_{\mu\nu\rho\sigma}{(-1)^g\over g}
   \Omega_{\nu\rho\sigma}^{[a_0\cdots a_g]}(n+\widehat\mu).
\eqn\sixxthirteen
$$
Then by using the noncommutative rule~\fourxfour, it is easy to see
that
$$
   {\cal A}\,d^4x(d\theta)^g\simeq
   \sum_n\Bigl[
   \sym(A^{a_0}C^{a_1}\cdots C^{a_g})\Omega^{[a_0\cdots a_g]}
   +C^{a_1}\cdots C^{a_g}\widetilde X^{[a_1\cdots a_g]}d^4x\Bigr].
\eqn\sixxfourteen
$$
On the other hand, the divergence-free condition~\sixxfive\ becomes
$$
   d\Omega^{[a_0\cdots a_g]}=0.
\eqn\sixxfifteen
$$
We can now apply the covariant Poincar\'e lemma~\fivexeleven\ to the
3-form~$\Omega^{[a_0\cdots a_g]}$ because it depends only on the field
strength. This yields
$$
   \Omega^{[a_0\cdots a_g]}=d\alpha_2^{[a_0\cdots a_g]}
   [\{F_i\}]+B_3^{[a_0\cdots a_g]}+F^bB_1^{[a_0\cdots a_g]b},
\eqn\sixxsixteen
$$
and the contribution of~$\alpha_2^{[a_0\cdots a_g]}$ can be absorbed
into the second term of~eq.~\sixxfourteen\ up to a trivial part,
because
$$
\eqalign{
   &\sum_n\sym(A^{a_0}C^{a_1}\cdots C^{a_g})
   d\alpha_2^{[a_0\cdots a_g]}
\cr
   &=\sum_n(-1)^g\sym(F^{a_0}C^{a_1}\cdots C^{a_g})
   \alpha_2^{[a_0\cdots a_g]}
\cr
   &\qquad
   +s\sum_n\biggl[(-1)^{g+1}{g\over2}
   \sym(A^{a_0}A^{a_1}C^{a_2}\cdots C^{a_g})\alpha_2^{[a_0\cdots a_g]}
   \biggr].
\cr
}
\eqn\sixxseventeen
$$
Similarly, from the covariant Poincar\'e lemma, we have
$$
\eqalign{
   &\widetilde X^{[a_1\cdots a_g]}d^4x
\cr
   &=d\alpha_3^{[a_1\cdots a_g]}[\{F_i\}]
   +{\cal L}^{[a_1\cdots a_g]}\,d^4x
   +B_4^{[a_1\cdots a_g]}+F^bB_2^{[a_1\cdots a_g]b}
   +F^bF^cB_0^{[a_1\cdots a_g](bc)},
\cr
}
\eqn\sixxeighteen
$$
and it is easy to see that $\alpha_3^{[a_1\cdots a_g]}$ does not
contribute to the nontrivial part.

So, up to this stage, we have obtained
$$
\eqalign{
   {\cal A}\,d^4x(d\theta)^g
   &\simeq\sum_n\Bigl[
   C^{a_1}\cdots C^{a_g}{\cal L}^{[a_1\cdots a_g]}\,d^4x
\cr
   &\qquad\quad+C^{a_1}\cdots C^{a_g}
   (B_4^{[a_1\cdots a_g]}+F^bB_2^{[a_1\cdots a_g]b}
   +F^bF^cB_0^{[a_1\cdots a_g](bc)})
\cr
   &\qquad\quad+\sym(A^{a_0}C^{a_1}\cdots C^{a_g})
   (B_3^{[a_0\cdots a_g]}+F^bB_1^{[a_0\cdots a_g]b})\Bigr],
\cr
}
\eqn\sixxnineteen
$$
where $\sum_n\delta{\cal L}^{[a_1\cdots a_g]}\neq0$ under a certain
local variation of the gauge potential. Formally this expression is
identical to the list of nontrivial solutions in the continuum theory
(see~eq.~(6.24) of the second reference of~ref.~[\BRA]). Recall
however that eq.~\sixxnineteen\ is an expression in the context of
noncommutative differential calculus and that it is valid for a finite
lattice spacing~$a\neq0$.

It is easy to see that eq.~\sixxnineteen\ satisfies
$\delta_B{\cal A}=0$. Does eq.~\sixxnineteen\ not contain BRS trivial
parts anymore? $\delta_B({\rm something})$ is always proportional to
a difference of the ghost field such as~$\Delta_\mu c^a$. However,
this does not necessarily imply that all terms
of~eq.~\sixxnineteen\ are BRS nontrivial. In contrast to the BRS
cohomology~\fivexten, this expression contains the summation~$\sum_n$.
Therefore, after ``integration by parts,'' a difference of ghost
fields may result.

In fact the term proportional to~$B_2$ contains BRS trivial parts.
Namely, by noting that $dC^a=-sA^a$, we have
$$
\eqalign{
   \sum_nC^{a_1}\cdots C^{a_g}F^bB_2^{[a_1\cdots a_g]b}
   &\simeq\sum_n\sym(C^{a_1}\cdots C^{a_g}dA^b)B_2^{[a_1\cdots a_g]b}
\cr
   &=\sum_n(-1)^gg\sym(sA^{a_1}C^{a_2}\cdots C^{a_g}A^b)
   B_2^{[a_1\cdots a_g]b}
\cr
   &\simeq-\sum_ng\sym(A^{a_1}C^{a_2}\cdots C^{a_g}sA^b)
   B_2^{[a_1\cdots a_g]b}
\cr
   &=\sum_n(-1)^gg\sym(sA^{b}C^{a_2}\cdots C^{a_g}A^{a_1})
   B_2^{[a_1\cdots a_g]b}.
\cr
}
\eqn\sixxtwenty
$$
Note that the commutator of~$C^a$ and the field strength 2-form~$F^b$
is proportional to a difference in the ghost field and thus, from the
BRS cohomology, it is BRS trivial. Therefore the ordering of~$C^a$
and~$F^b$ is arbitrary in the first expression of~eq.~\sixxtwenty\ up
to BRS trivial parts. We have used this fact for the first
$\simeq$~equality. By comparing the second line and the fourth line
of the above expression, we see that eq.~\sixxtwenty\ is equivalent to
$$
   \sum_nC^bC^{a_2}\cdots C^{a_g}F^{a_1}B_2^{[a_1\cdots a_g]b}
   =\sum_nC^{a_1}\cdots C^{a_g}F^bB_2^{[ba_2\cdots a_g]a_1}.
\eqn\sixxtwentyone
$$
A comparison with the left hand side of~eq.~\sixxtwenty\ shows that
the antisymmetric part of~$B_2^{[a_1\cdots a_g]b}$
under~$a_1\leftrightarrow b$ is BRS trivial~$\simeq0$. Therefore
$B_2^{[a_1\cdots a_g]b}$~must be symmetric
under~$a_1\leftrightarrow b$ to contribute nontrivial solutions.

Similarly, we have (suppressing $\sum_n$)
$$
\eqalign{
   C^{a_1}\cdots C^{a_g}F^bF^cB_0^{[a_1\cdots a_g](bc)}
   &\simeq\sym(C^{a_1}\cdots C^{a_g}F^b)F^cB_0^{[a_1\cdots a_g](bc)}
\cr
   &\simeq C^bC^{a_2}\cdots C^{a_g}F^{a_1}F^c
   B_0^{[a_1\cdots a_g](bc)},
\cr
}
\eqn\sixxtwentytwo
$$
and therefore $B_0^{[a_1\cdots a_g](bc)}$~must be symmetric under
$a_1\leftrightarrow b$.

The term proportional to~$B_1$ might also contain BRS trivial parts
depending on symmetry of indices. However, the noncommutativity
prevented us to imitate the procedure in the continuum theory~[\BRA].

Let us summarize the result: The general structure of local solutions
to the consistency condition~eq.~\onexthree\ is given
by~eq.~\sixxnineteen. The constant forms $B_2$ and~$B_0$ have the
following symmetries:
$$
   B_2^{[a_1\cdots a_g]b}=B_2^{[ba_2\cdots a_g]a_1},\qquad
   B_0^{[a_1\cdots a_g](bc)}=B_0^{[ba_2\cdots a_g](a_1c)}.
\eqn\sixxtwentythree
$$
The solution~\sixxnineteen\ {\it is\/} nontrivial, i.e., it cannot be
written as ${\cal A}=\delta_B{\cal B}$ by using a {\it local\/}
functional~${\cal B}$. The classical continuum limit
of~eq.~\sixxnineteen\ with~eq.~\sixxtwentythree\ coincides with the
nontrivial solutions in the continuum theory~[\BRA] (with a partial
exception for~$B_1^{[a_0\cdots a_g]b}$ mentioned above). Then if
eq.~\sixxnineteen~was BRS trivial, the classical continuum limit of
the local functional~${\cal B}$ would act as a counter term for the
nontrivial solutions in the continuum theory. But this contradicts
with the result of~ref.~[\BRA].

In this subsection, we have obtained the general nontrivial local
solutions with an arbitrary ghost number. For discussions of the gauge
anomaly in the next subsection, knowledge of solutions with ghost
number one is enough. The solutions with higher ghost number,\foot{%
The information about these is important~[\BRA] when one explicitly
considers the higher order sequence~${\cal A}_\ell$ with~$\ell\geq 4$
for the nonabelian anomaly; we do not pursue this in this paper.}
however, might become relevant for future applications. For example,
it might be possible to address the commutator anomaly~[\FAD] in the
context of lattice gauge theory starting with the above expressions.

\section{Gauge anomaly in abelian theory}
If we restrict solutions with the ghost number unity,
eq.~\sixxnineteen\ tells us that
$$
\eqalign{
   {\cal A}&\simeq\sum_n\Bigl\{
   c^a(n){\cal L}^a(n)
\cr
   &\qquad\quad+c^a(n)\Bigl[
   \alpha^a+\beta_{[\mu\nu]}^{(ab)}F_{\mu\nu}^b(n)
   +\gamma^{(abc)}\varepsilon_{\mu\nu\rho\sigma}
   F_{\mu\nu}^b(n)F_{\rho\sigma}^c(n+\widehat\mu+\widehat\nu)\Bigr]
\cr
   &\qquad\quad+\Bigl[A_\mu^a(n)c^b(n+\widehat\mu)
   +c^b(n)A_\mu^a(n)\Bigr]
   \Bigl[f_\mu^{[ab]}
   +g_{[\mu\nu\rho]}^{[ab]c}F_{\nu\rho}^c(n+\widehat\mu)\Bigr]\Bigr\},
\cr
}
\eqn\sixxtwentyfour
$$
where we have used the noncommutative rule~\fourxfour\ and the
symmetry of indices~\sixxtwentythree. In this expression, the function
${\cal L}^a(n)$ satisfies $\sum_n\delta{\cal L}^a(n)\neq0$ for a
certain local variation.

Eq.~\sixxtwentyfour\ provides the general candidate of nontrivial
local gauge anomalies in the abelian theory~$G=U(1)^N$. However,
depending on the situation, we may further restrict the coefficients
in various ways.

\noindent
(1)~When $G=U(1)$, the last line vanishes due to the
anti-symmetrization of indices. Eq.~\sixxtwentyfour\ then reproduces
L\"uscher's result~\onexone\ except for the ``non-topological
term''~${\cal L}^a$.

\noindent
(2)~The non-topological term~${\cal L}^a$ and the term proportional
to~$f_\mu^{[ab]}$ do not appear, if the anomaly has the topological
property~$\delta{\cal A}[c,A]=0$ for~$c^a(n)\to{\rm const.}$, where
$\delta$~is an arbitrary local variation of the gauge potential.

\noindent
(3)~If the couplings of the Weyl fermion to gauge fields have the same
structure for all $U(1)$ factors except coupling constants
(practically this is always the case), then all the coefficients are
independent of group indices and we have $\alpha^a\to\alpha$,
$\beta_{[\mu\nu]}^{(ab)}\to\beta_{[\mu\nu]}$, $\gamma^{(abc)}\to%
\gamma$, $f_\mu^{[ab]}\to0$, $g_{[\mu\nu\rho]}^{[ab]c}\to0$.

\noindent
(4)~From the dimension counting, all the terms except~${\cal L}^a$ and
the term proportional to~$\gamma^{(abc)}$ have negative powers of the
lattice spacing as the overall coefficient. Therefore, if the
classical continuum limit~$a\to0$ of~${\cal A}$ is finite (for a
smooth background), all the terms except~${\cal L}^a$
and~$\gamma^{(abc)}$ must be absent. In particular, if
$\lim_{a\to0}{\cal A}$ reproduces the gauge anomaly in the continuum
theory, then $\gamma^{(abc)}=-\epsilon_H/(96\pi^2)$ for a single Weyl
fermion.

Let us assume that (2) and~(4) hold. Then we have a content of the
theorem for the abelian gauge theory which we stated in~section~3.

\chapter{Nonabelian extension}
In this section, we study the gauge anomaly for a general (compact)
gauge group~$G=\prod_\alpha G_\alpha$, where $G_\alpha$~is a simple
group or a $U(1)$ factor. The candidate of the anomaly is given by the
solution with ghost number one to the consistency
condition~\onexthree. The general solution to~eq.~\onexthree\ is
expressed as
$$
   {\cal A}=\widetilde{\cal A}+\delta_B{\cal B},
\eqn\sevenxone
$$
where $\widetilde{\cal A}\neq\delta_B{\cal B}$ is the BRS nontrivial
part (${\cal B}$ is a local functional). The BRS transformation is
given by eq.~\onextwo\ and it takes the following form in terms of the
gauge potential
$$
   \delta_BA_\mu(n)
   ={1\over2}A_\mu(n)\wedge
   \biggl[\coth{1\over2}A_\mu(n)\wedge\Delta_\mu c(n)
   +c(n)+c(n+\widehat\mu)\biggr],
\eqn\sevenxtwo
$$
where $X\wedge Y=[X,Y]$, $X^2\wedge Y=[X,[X,Y]]$ and so on,
and~$1\wedge Y=Y$ is understood. We shall use both the matrix
notation~$A_\mu(n)$ and~$c(n)$, and the component
notation~$A_\mu(n)=\sum_sA_\mu^a(n)T^a$ and~$c(n)=\sum_ac^a(n)T^a$.

To make our problem tractable, we make the following assumptions about
the anomaly~${\cal A}$.

\noindent
(I)~${\cal A}$ is a smooth and local functional of the gauge
potential~$A_\mu$ and the ghost field~$c$.

\noindent
(II)~The classical continuum limit of~${\cal A}$ reproduces the
anomaly in the continuum theory as in~eq.~\twoxtwentyeight.

\noindent
(III)~$U(1)$ gauge anomalies in~${\cal A}$ have the topological
property as in~eq.~\addsix.

Under these assumptions, in section~7.2, we show that the
anomaly~${\cal A}$, if it exists, is {\it unique\/} (up to the BRS
trivial part) to all orders of the gauge potential. The unique anomaly
is proportional to the gauge anomaly in the continuum theory and this
establishes the theorem for nonabelian theories, stated in section~3.
In section~7.3, we show that such a solution in fact {\it exists}. As
a preparation for sec.~7.2, we need the following lemma.

\section{Basic lemma: Adjoint invariance}
\proclaim Adjoint invariance.
Without loss of generality, one can assume that a nontrivial local
solution~$\widetilde{\cal A}$ is invariant under the adjoint
transformation
$$
   \delta^a\widetilde{\cal A}=0,
\eqn\sevenxthree
$$
where the adjoint transformation~$\delta^a$ is defined by
$$
   \delta^aU(n,\mu)=[T^a,U(n,\mu)],\qquad
   \delta^aA_\mu^b(n)=-if^{abc}A_\mu^c(n),\qquad
   \delta^ac^b(n)=-if^{abc}c^c(n).
\eqn\sevenxfour
$$

\noindent
{\it Note}. There is freedom to add a BRS trivial
part~$\delta_B{\cal B}$ to a nontrivial solution. The above lemma
asserts that it is always possible to choose ${\cal B}$ such that
$\widetilde{\cal A}$~is adjoint invariant. The adjoint
transformation~$\delta^a$ satisfies the following relations
$$
   [\delta_B,\delta^a]=0,\qquad
   [\Delta_\mu^*,\delta^a]=0,\qquad
   [\delta^a,\delta^b]=if^{abc}\delta^c.
\eqn\sevenxfive
$$

\noindent
{\it Proof}. The functional~$\widetilde{\cal A}$ is local, i.e., the
field~$\widetilde a(n)$ in~$\widetilde{\cal A}=\sum_n\widetilde a(n)$
is a local field. We express~$\widetilde a(n)$ in terms of the
following set of variables, which was introduced in the proof of the
abelian BRS cohomology in~ref.~[\FUJ]:
$$
   A_i^a=(\Delta_1)^{p_1}\cdots(\Delta_\mu)^{p_\mu}A_\mu^a(n),\qquad
   F_i^a=(\Delta_1)^{p_1}\cdots(\Delta_D)^{p_D}F_{\mu\nu}^a(n),
\eqn\sevenxsix
$$
for the gauge potential ($D$ is the dimension of the lattice) and
$$
   c_i^a=\delta_0A_i^a
   =(\Delta_1)^{p_1}\cdots(\Delta_\mu)^{p_\mu}\Delta_\mu c^a(n),
   \qquad{\rm and}\qquad c^a(n),
\eqn\sevenxseven
$$
for the ghost field; here $\delta_0$~is the {\it abelian\/} BRS
transformation, $\delta_0A_\mu^a(n)=\Delta_\mu c^a(n)$
and~$\delta_0c^a(n)=0$. In these expressions, the
symbol~$(\Delta_\mu)^p$ ($p$~is an integer) has been defined by
$$
   (\Delta_\mu)^p=\cases{\Delta_\mu^p,&for $p>0$,\cr
                         1,&for $p=0$,\cr
                         \Delta_\mu^{*-p},&for $p<0$.\cr}
\eqn\sevenxeight
$$
Then it can be shown~[\FUJ] that these variables, $A_i^a$, $F_i^a$,
$c_i^a$ and~$c^a(n)$ span a (over)complete set, i.e., the
field~$\widetilde a(n)$ can be expressed as a function of these
variables. A little thought shows that the relation
$$
   \biggl[{\partial\over\partial c^a(n)},\Delta_\mu^*\biggr]=0,
\eqn\sevenxnine
$$
holds for arbitrary functions of these variables.

Since the {\it nonabelian\/} BRS
transformation~\onextwo\ or~\sevenxtwo\ has the structure,
$$
\eqalign{
   &\delta_BA_\mu^a(n)=if^{abc}A_\mu^b(n)c^c(n)
   +(\hbox{terms proportional to $\Delta_\mu c^a$}),
\cr
   &\delta_Bc^a(n)=-{1\over2}if^{abc}c^b(n)c^c(n),
\cr
}
\eqn\sevenxten
$$
we have
$$
   \delta_B\left\{\matrix{A_i^a\cr F_i^a\cr c_i^a\cr}\right\}
   =-c^b(n)\delta^b\left\{\matrix{A_i^a\cr F_i^a\cr c_i^a\cr}\right\}
   +(\hbox{terms proportional to $c_i^a$}),
\eqn\sevenxeleven
$$
and thus
$$
   \delta^a
   =-\biggl\{\delta_B,{\partial\over\partial c^a(n)}\biggr\},
\eqn\sevenxtwelve
$$
on functions of the variables $A_i^a$, $F_i^a$, $c_i^a$ and~$c^a(n)$
(for $c^a(n)$ this follows from~eq.~\sevenxten).

The remaining argument to prove the lemma is almost identical to that
of~ref.~[\BRA]. We introduce the Casimir operator:
$$
   {\cal O}_K=g^{a_1\cdots a_{m(K)}}
   \delta^{a_1}\cdots\delta^{a_{m(K)}},
\eqn\sevenxthirteen
$$
where~$g^{a_1\cdots a_{m(K)}}=\str T^{a_1}\cdots T^{a_{m(K)}}$ are
totally symmetric constants and $K$~runs from~1 to the rank of the
semisimple part of the group~$G$~[\ORA]. Using the completeness of
eigenfunctions of~${\cal O}_K$, we decompose~$\widetilde a(n)$
according to the representation~$\lambda$, $\widetilde a(n)=%
\sum_\lambda\widetilde a^\lambda(n)$, where
$$
   {\cal O}_K\widetilde a^\lambda(n)
   =k(K,\lambda)\widetilde a^\lambda(n),
\eqn\sevenxfourteen
$$
and $k(K,\lambda)$ is the eigenvalue. Since
$\delta_B\widetilde{\cal A}=\sum_n\delta_B\widetilde a(n)=0$, the dual
of the algebraic Poincar\'e lemma~\fivexone\ ($p=D$) shows that
$$
   \delta_B\widetilde a(n)=\Delta_\mu^*X_\mu(n),
\eqn\sevenxfifteen
$$
where $X_\mu(n)$ is a local field. We again apply to this equation the
decomposition similar to eq.~\sevenxfourteen:
$$
   \delta_B\widetilde a^\lambda(n)=\Delta_\mu^*X_\mu^\lambda(n),
\eqn\sevenxsixteen
$$
where use of relations~\sevenxfive\ has been made.

Now suppose that there exist $K$ and~$\lambda$ such that
$k(K,\lambda)\neq0$ in~eq.~\sevenxfourteen. Then, by using
eqs.~\sevenxthirteen, \sevenxtwelve, \sevenxfive\ and~\sevenxsixteen,
we have
$$
\eqalign{
   \widetilde a^\lambda(n)
   &=\delta_B{-1\over k(K,\lambda)}
   g^{a_1\cdots a_{m(K)}}\delta^{a_{m(K)}}\cdots\delta^{a_2}
   {\partial\over\partial c^{a_1}(n)}\widetilde a^\lambda(n)
\cr
   &\qquad+\Delta_\mu^*{-1\over k(K,\lambda)}
   g^{a_1\cdots a_{m(K)}}\delta^{a_{m(K)}}\cdots\delta^{a_2}
   {\partial\over\partial c^{a_1}(n)}X_\mu^\lambda(n).
\cr
}
\eqn\sevenxseventeen
$$
Namely, $\widetilde{\cal A}$ contains a BRS trivial
part~$\sum_n\widetilde a^\lambda(n)$ which we can remove
by~$\delta_B{\cal B}$. After repeating this procedure, all the
eigenvalues~$k(K,\lambda)$ in~eq.~\sevenxfourteen\ are made to vanish
and this implies that $\lambda$~is the singlet representation.
Therefore, we can always assume that a BRS nontrivial solution is
adjoint invariant, $\delta^a\widetilde{\cal A}=0$.\QED

We next derive a constraint for~$\widetilde{\cal A}$ following from
the assumption~(III) made above and the lemma~\sevenxthree. Set
$c^a(n)\to c^a={\rm const}$. Since the ghost number
of~$\widetilde{\cal A}$ is unity, we can write it as
$$
   \widetilde{\cal A}=c^ak^{a\lambda}X^\lambda[A],
\eqn\sevenxeighteen
$$
where $k^{a\lambda}$ are constants and $\lambda$ labels the linearly
independent functional~$X^\lambda[A]$. We then consider the following
two cases separately:

\noindent
(1) When the index~$a$ of the ghost field in
eq.~\sevenxeighteen\ belongs to a $U(1)$~factor group~$U(1)_\alpha$,
we have
$$
   \widetilde{\cal A}=c^{U(1)_\alpha}X[A].
\eqn\sevenxnineteen
$$
However, from assumption~(III), we have~$\delta X=0$,
where~$\delta$~is an arbitrary local variation of the gauge
potential.\foot{Note that the addition of~$\delta_B{\cal B}$ with a
{\it local\/} term ${\cal B}$ does not influence the topological
property, as noted in~sec.~2.} 

\noindent
(2)~When the index~$a$ of the ghost field in
eq.~\sevenxeighteen\ belongs to a simple group, the BRS
transformation~\onextwo\ or~\sevenxtwo\ becomes for $c^a(x)\to%
c^a={\rm const.}$,
$$
\eqalign{
   &\delta_BA_\mu^a(n)=-if^{abc}c^bA_\mu^c(n)
   =-c^b\delta^bA_\mu^a(n),
\cr
   &\delta_Bc^a=-{1\over2}if^{abc}c^bc^c
   =-\delta_Bc^a-c^b\delta^bc^a,
\cr
}
\eqn\sevenxtwenty
$$
where $\delta^a$ is the adjoint transformation. But since the
lemma~\sevenxthree\ asserts that $\delta^a\widetilde{\cal A}=0$, the
consistency condition becomes for $c^a(n)\to{\rm const.}$,
$$
   \delta_B\widetilde{\cal A}
   =-\delta_B(c^ak^{a\lambda})X^\lambda[A]=0.
\eqn\sevenxtwentyone
$$
This requires $\delta_B(c^ak^{a\lambda})=0$. Then the Lie algebra
cohomology in~ref.~[\BRA] asserts that $c^ak^{a\lambda}=\tr c=0$ for a
simple group. This shows that $\widetilde{\cal A}=0$
for~$c^a(n)\to{\rm const}$. This conclusion might be dangerous because
there is a possibility of having a total divergence. To avoid this, it
is enough to consider a local variation which implies
$\delta X^\lambda=0$ where $\delta$~is an arbitrary local variation of
the gauge potential.

Combining above (1) and~(2), we see that the assumption~(III) implies
$$
   \delta\widetilde{\cal A}=0,
   \qquad{\rm for}\quad c^a(n)\to{\rm const.},
\eqn\sevenxtwentytwo
$$
where $\delta$~is an arbitrary local variation of the gauge potential.
This provides a strong constraint for the possible form
of~$\widetilde{\cal A}$, as will be seen in the next subsection.

\section{Uniqueness of the nontrivial anomaly}
We now expand the anomaly~${\cal A}$~\sevenxone\ in powers of the
gauge potential as
$$
   {\cal A}=\sum_{\ell=1}^\infty{\cal A}_\ell,\qquad
   \widetilde{\cal A}=\sum_{\ell=1}^\infty\widetilde{\cal A}_\ell,
   \qquad{\cal B}=\sum_{\ell=1}^\infty{\cal B}_\ell,
\eqn\sevenxtwentythree
$$
where $\ell$~stands for the number of powers of $c$ and~$A_\mu$
(recall that the ghost number of ${\cal A}$ is unity). We decompose
also the BRS transformation~\sevenxtwo\ according to powers of the
fields, $\delta_B=\sum_{\ell=0}^\infty\delta_\ell$, where
$$
\eqalign{
   &\delta_0A_\mu(n)=\Delta_\mu c(n),\qquad
   \delta_0c(n)=0,
\cr
   &\delta_1A_\mu(n)
   ={1\over2}\bigl[A_\mu(n),c(n)+c(n+\widehat\mu)\bigr],\qquad
   \delta_1c(n)=-c(n)^2,
\cr
   &\delta_{2k}A_\mu(n)
   =(-1)^{k-1}{B_k\over(2k)!}
   \bigl[\underbrace{
   A_\mu(n),\bigl[A_\mu(n),\cdots,\bigl[A_\mu(n)
   }_{2k},\Delta_\mu c(n)\bigr]\cdots\bigr]\bigr],
\cr
   &\delta_{2k}c(n)=0,\qquad{\rm for}\quad k\geq1.
\cr
}
\eqn\sevenxtwentyfour
$$
Here $B_k$~is the Bernoulli number, and $\delta_{2k+1}=0$
for~$k\geq1$. Note that, in terms of components $A_\mu^a$ and~$c^a$,
$\delta_0$ has an identical form to the abelian BRS
transformation~\fourxseven. The nilpotency~$\delta_B^2=0$ implies
$$
   \sum_{k=0}^\ell\delta_k\delta_{\ell-k}=0,\qquad{\rm for}\quad
   \ell\geq0,
\eqn\sevenxtwentyfive
$$
and the consistency condition~\onexthree\ takes the form
$$
   \delta_0\widetilde{\cal A}_\ell
   =-\sum_{k=1}^{\ell-1}\delta_k\widetilde{\cal A}_{\ell-k},
   \qquad{\rm for}\quad\ell\geq1.
\eqn\sevenxtwentysix
$$
Since
$$
   {\cal A}_\ell=\widetilde{\cal A}_\ell
   +\sum_{k=0}^{\ell-1}\delta_k{\cal B}_{\ell-k},
\eqn\sevenxtwentyseven
$$
if $\widetilde{\cal A}_\ell$ contains $\delta_0$-trivial
part~$\delta_0{\cal B}_\ell'$, ${\cal B}_\ell'$ can always be absorbed
into ${\cal B}_\ell$. Therefore we can assume that
$$
   \hbox{$\delta_0{\cal B}_\ell'$ in $\widetilde{\cal A}_\ell$
   can always be neglected}.
\eqn\sevenxtwentyeight
$$

The constraint~\sevenxtwentytwo\ has to hold for each order:
$$
   \delta\widetilde{\cal A}_\ell=0,\qquad{\rm for}\quad
   c^a(n)\to{\rm const.},
\eqn\sevenxtwentynine
$$
where $\delta$~is an arbitrary local variation of the gauge potential.
Also the correct classical continuum limit~(II) requires
$$
   \widetilde{\cal A}_\ell\buildrel{a\to0}\over\rightarrow
   \hbox{$O(A^{\ell-1})$ term of eq.~\twoxtwentyeight}.
\eqn\sevenxthirty
$$

We consider the local solution to the consistency
condition~\sevenxtwentysix\ which satisfies the
conditions~\sevenxtwentynine\ and~\sevenxthirty, order by order. The
first equation in~eq.~\sevenxtwentysix\ is
$$
   \delta_0\widetilde{\cal A}_1=0.
\eqn\sevenxthirtyone
$$
This equation is completely identical to the consistency condition in
the abelian theory. Therefore, from our result in the preceding
section, eq.~\sixxtwentyfour, the general form
of~$\widetilde{\cal A}_1$ which satisfies
eqs.~\sevenxtwentynine\ and~\sevenxthirty\ is given by
$$
   \widetilde{\cal A}_1=0,
\eqn\sevenxthirtytwo
$$
where we have noted eq.~\sevenxtwentyeight. The next equation
in~eq.~\sevenxtwentysix\ is
$$
   \delta_0\widetilde{\cal A}_2=0.
\eqn\sevenxthirtythree
$$
Again from the result in the abelian theory~\sixxtwentyfour, we have
$$
   \widetilde{\cal A}_2=0,
\eqn\sevenxthirtyfour
$$
where use of eqs.~\sevenxtwentyeight--\sevenxthirty\ has been made to
conclude this.

The solution to the next equation
$$
   \delta_0\widetilde{\cal A}_3=0,
\eqn\sevenxthirtyfive
$$
has a $\delta_0$-nontrivial part. From eq.~\sixxtwentyfour, and from
eqs.~\sevenxtwentyeight--\sevenxthirty, we have
$$
\eqalign{
   \widetilde{\cal A}_3&=-{\epsilon_H\over24\pi^2}\sum_n
   \Bigl\{\varepsilon_{\mu\nu\rho\sigma}{1\over2}\tr c^{(\alpha)}(n)
   \Delta_\mu\Bigl\{A_\nu^{(\alpha)}(n),
   \Delta_\rho A_\sigma^{(\alpha)}(n+\widehat\nu)\Bigr\}
\cr
   &\qquad\qquad\qquad\quad
   +\varepsilon_{\mu\nu\rho\sigma}c^{U(1)_\beta}(n)
   \Delta_\mu\Bigl[A_\nu^{U(1)_\beta}(n)
   \Delta_\rho A_\sigma^{U(1)_\beta}(n+\widehat\nu)\Bigr]
\cr
   &\qquad\qquad\qquad\quad
   +\varepsilon_{\mu\nu\rho\sigma}c^{U(1)_\beta}(n)
   \tr\Delta_\mu\Bigl[A_\nu^{(\alpha)}(n)
   \Delta_\rho A_\sigma^{(\alpha)}(n+\widehat\nu)\Bigr]
\cr
   &\qquad\qquad\qquad\quad
   +\varepsilon_{\mu\nu\rho\sigma}\tr c^{(\alpha)}(n)
   \Delta_\mu\Bigl[A_\nu^{U(1)_\beta}(n)
   \Delta_\rho A_\sigma^{(\alpha)}(n+\widehat\nu)\Bigr]
\cr
   &\qquad\qquad\qquad\quad
   +\varepsilon_{\mu\nu\rho\sigma}\tr c^{(\alpha)}(n)
   \Delta_\mu\Bigl[A_\nu^{(\alpha)}(n)
   \Delta_\rho A_\sigma^{U(1)_\beta}(n+\widehat\nu)\Bigr]\Bigr\},
\cr
}
\eqn\sevenxthirtysix
$$
where we have used relation~\fivexfiftyeight\ to make the
property~\sevenxtwentynine\ manifest. Note that the
condition~\sevenxtwentynine\ is crucial to eliminate the possibility
that the ${\cal L}^a$~term in~eq.~\sixxtwentyfour\ appears.

The next equation in~eq.~\sevenxtwentysix\ is
$$
   \delta_0\widetilde{\cal A}_4=-\delta_1\widetilde{\cal A}_3.
\eqn\sevenxthirtyseven
$$
The solution to this equation~$\widetilde{\cal A}_4$, if it exists, is
{\it unique}. If another $\widetilde{\cal A}_4'$ which also satisfies
the conditions~\sevenxtwentynine\ and~\sevenxthirty\ exists, then
$$
   \delta_0(\widetilde{\cal A}_4'-\widetilde{\cal A}_4)=0,
\eqn\sevenxthirtyeight
$$
and the quantity inside the brackets again satisfies
eq.~\sevenxtwentynine.
Eqs.~\sevenxtwentyeight\ and~\sixxtwentyfour\ then imply that
$\widetilde{\cal A}_4'-\widetilde{\cal A}_4=0$.

The above argument can be repeated for higher
$\widetilde{\cal A}_\ell$'s. Suppose that a sequence for the
nontrivial part, $\widetilde{\cal A}_3$, $\widetilde{\cal A}_4$,
\dots, $\widetilde{\cal A}_{\ell-1}$, has been obtained. Then the next
term~$\widetilde{\cal A}_\ell$ has to satisfy
eq.~\sevenxtwentysix\ and the
conditions~\sevenxtwentynine\ and~\sevenxthirty. Then the same
argument as above shows that the solution~$\widetilde{\cal A}_\ell$,
if it exists, is {\it unique}.

We have seen that the sequence~$\widetilde{\cal A}_\ell$ for the
nontrivial anomaly~$\widetilde{\cal A}$, which satisfies the
assumptions~(I), (II) and~(III), if it exists, is unique up to a BRS
trivial part. There is no free parameter which can appear in
higher~$\widetilde{\cal A}_\ell$'s. Moreover, this uniqueness shows
that the anomaly cancellation in the continuum theory implies that of
the lattice theory: If the first nontrivial
term~$\widetilde{\cal A}_3$~\sevenxthirtysix, which is proportional to
the anomaly in the continuum theory~$\tr_{R-L}T^a\{T^b,T^c\}$ etc., is
canceled among the fermion multiplet, then the subsequent sequence
of~$\widetilde{\cal A}_\ell$ for~$\ell\geq4$ is completely canceled.
In other words, the possible nontrivial local anomaly on the lattice,
$\widetilde{\cal A}$, under the assumptions~(I)--(III), is always
proportional to the anomaly in the continuum theory, to all orders in
powers of the gauge potential. This establishes our theorem for
nonabelian theories, stated in section~3.

The existence of the nontrivial sequence $\widetilde{\cal A}_\ell$
for~$\ell\geq4$ might be examined by repeatedly solving
eq.~\sevenxtwentysix. However, eq.~\sevenxtwentyfour\ suggests that
the explicit form of higher~$\widetilde{\cal A}_\ell$'s will become
quite complicated as $\ell$~increases. In the next subsection, instead
of this analysis, we will give a ``compact'' form of the nontrivial
solution which manifestly satisfies eq.~\sevenxtwentytwo\ and the
assumptions~(I) (at least for the perturbative region~\twoxnine)
and~(II). This explicitly shows the existence of the nontrivial
sequence, $\widetilde{\cal A}_\ell$ with~$\ell\geq4$. To write down
the compact solution, however, we need the interpolation technique of
lattice fields with which the BRS transformation takes a quite simple
form. Therefore, we give a quick summary of the method of~ref.~[\GOC]
in the first part of the next subsection.

\section{Compact form of the nontrivial anomaly}
First we recapitulate the essence of the interpolation method of
lattice fields in ref.~[\GOC] (simply extended for infinite
lattices).\foot{%
Under the same conditions we will assume, the method of ref.~[\HER]
might be adopted as well.} For our argument, the interpolation method
has to possess several properties which we will verify. To distinguish
{}from the fields defined on lattice sites~$n$, we use the continuous
coordinate~$x$ for interpolated fields.

The method of ref.~[\GOC] consists of the following two steps.

\noindent
Step~1. One first constructs the interpolated gauge
potential~$A_\mu^{(m)}(x)$ within each hypercube~$h(m)$, here
$m$~stands for the origin of the hypercube, such that the gauge
potentials in neighboring hypercubes $h(m-\widehat\mu)$ and~$h(m)$ are
related by the transition function~$v_{m,\mu}(x)$ of L\"uscher's
principal fiber bundle~[\LUSCHER],
$$
   A_\lambda^{(m-\widehat\mu)}(x)
   =v_{m,\mu}(x)\Bigl[
   \partial_\lambda+A_\lambda^{(m)}(x)\Bigr]v_{m,\mu}(x)^{-1},
\eqn\sevenxthirtynine
$$
on the intersection of the two hypercubes
$x\in h(m-\widehat\mu)\cap h(m)$ (which is a 3-dimensional cube). For
the transition function~$v_{m,\mu}(x)$ to be well-defined, the gauge
field configuration must be ``non-exceptional''~[\LUSCHER]. As already
noted, it can be shown that if $\epsilon$ in~eq.~\twoxone\ is
sufficiently small, the gauge field configuration is non-exceptional.
So we assume that $\epsilon$~has been chosen so that this is the case.
The gauge potential which satisfies eq.~\sevenxthirtynine\ can be
constructed, starting with a special gauge $A_\lambda^{(m)}(x)=0$
at~$x\sim m$. The explicit expression of~$A_\lambda^{(m)}(x)$ in terms
of~$v_{m,\mu}(x)$, which is eventually expressed by the link
variables~$U$~[\LUSCHER], is given in~ref.~[\GOC]. We do not reproduce
it here because it is rather involved and we do not need the explicit
form in what follows. The interesting property of~$A_\mu^{(m)}(x)$
is~[\GOC]
$$
   {\cal P}\exp\biggl[\int\nolimits_0^1dt\,A_\mu^{(m)}
   (n+(1-t)\widehat\mu)\biggr]=u_{n,n+\widehat\mu}^m.
\eqn\sevenxforty
$$
Namely, the Wilson line constructed from the interpolated gauge
potential~$A_\mu^{(m)}(x)$ coincides with the link variable in the
complete axial gauge of ref.~[\LUSCHER].

\noindent
Step~2.1. The section of the principal fiber bundle~[\LUSCHER]
$$
   w^m(n)=U(m,1)^{z_1}U(m+z_1\widehat1,2)^{z_2}
   U(m+z_1\widehat1+z_2\widehat2,3)^{z_3}
   U(m+z_1\widehat1+z_2\widehat2+z_3\widehat3,4)^{z_4}
   \in G,
\eqn\sevenxfortyone
$$
is defined for each lattice site~$n$ belonging to the
hypercube~$h(m)$ where $n=m+\sum_\mu z_\mu\widehat\mu$. This section
is then smoothly interpolated, first on the links, next on the
plaquettes, on the cubes, and finally inside the hypercube $h(m)$. At
this stage, if the homotopy group~$\pi_{M-1}(G)$ is nontrivial, the
smooth interpolation of the section~$w^m(x)$ into a $M$-dimensional
(sub)lattice may fail, depending on the configuration of the
section~$w^m(x)$ on a boundary of the $M$-dimensional (sub)lattice.
For example, for~$G=U(1)$, $\pi_1(U(1))=Z$, and if the local winding
of~$w^m(x)$ around a boundary of the plaquette $p(m,\mu,\nu)$,
$$
   Q(m,\mu,\nu)={i\over2\pi}
   \int_{\partial p(m,\mu,\nu)}dx_\mu\,\varepsilon_{\mu\nu}\,
   w^m(x)^{-1}\partial_\nu w^m(x),
\eqn\sevenxfortytwo
$$
does not vanish, then the interpolation of the section~$w^m(x)$ into
the plaquette~$p(m,\mu,\nu)$ develops a singularity. Similarly,
for~$G=SU(2)$, $\pi_3(SU(2))=Z$, and the local winding is given
by\foot{The total winding $Q=\sum_mQ(m)$ on a finite periodic lattice
is nothing but L\"uscher's topological charge~[\LUSCHER].}
$$
   Q(m)={1\over24\pi^2}\int_{\partial h(m)}d^3x_\mu\,
   \varepsilon_{\mu\nu\rho\sigma}\tr
   w^m(x)^{-1}\partial_\nu w^m(x)w^m(x)^{-1}\partial_\rho w^m(x)
   w^m(x)^{-1}\partial_\sigma w^m(x).
\eqn\sevenxfortythree
$$
If $Q(m)$~does not vanish, then the interpolation of~$w^m(x)$ into
the hypercube~$h(m)$ develops the singularity. If these singularities
arise, the description in term of the interpolated fields becomes
inadequate.\foot{%
The procedure of~ref.~[\HER] can avoid this difficulty.} Fortunately,
all local windings vanish within the perturbative region~\twoxnine,
for sufficiently small~$\epsilon$. If $\epsilon$~in~\twoxnine\ is
sufficiently small, the norm of the exponent of a product of several
link variables is also small, and the expression of the interpolated
section~[\GOC] cannot have the ``jump'' on a boundary of the
$M$-dimensional (sub)lattice. This implies that there is no local
winding.

\noindent
Step~2.2. With the smooth interpolated section~$w^m(x)$, we define
the ``global'' interpolated gauge potential by
$$
   A_\lambda(x)
   =w^m(x)^{-1}\Bigl[
   \partial_\lambda+A_\lambda^{(m)}(x)\Bigr]w^m(x),
\eqn\sevenxfortyfour
$$
for $x\in h(m)$. The resulting interpolated gauge
potential~$A_\lambda(x)$ is Lie algebra valued.

Now, we need the following properties of the interpolation method to
express the nontrivial local solution.

\noindent
(i)~The gauge covariance. This is the most important property for our
purpose. Namely, there exists a smooth interpolation of the gauge
transformation (in our present context this becomes a smooth
interpolation of the ghost field) and the lattice gauge (BRS)
transformation on the link variables takes an {\it identical\/} form
as that of the continuum theory. This property was shown
in~ref.~[\GOC]. Therefore, the BRS transformation~\onextwo\ induces
$$
   \delta_BA_\mu^a(x)=\partial_\mu c^a(x)+if^{abc}A_\mu^b(x)c^c(x),
   \qquad
   \delta_Bc^a(x)=-{1\over2}if^{abc}c^b(x)c^c(x),
\eqn\sevenxfortyfive
$$
on the {\it interpolated fields}.

\noindent
(ii)~The transverse continuity. This means that the gauge
potential~$A_\lambda(x)$ is continuous inside each hypercube and, on
the intersection of two neighboring hypercubes
$x\in h(m-\widehat\mu)\cap h(m)$, the component {\it transverse\/} to
this intersection (namely, $\lambda\neq\mu$) is continuous across this
intersection. We need this property because otherwise boundary terms
arising from integration by parts are not cancelled in the following
expression. It is easy to see this property if one notes that
L\"uscher's transition function~[\LUSCHER] and the interpolated
section~[\GOC] are related by
$$
   v_{m,\mu}(x)=w^{m-\widehat\mu}(x)w^m(x)^{-1},\qquad
   {\rm for}\quad x\in h(m-\widehat\mu)\cap h(m).
\eqn\sevenxfortysix
$$
Then from eqs.~\sevenxthirtynine\ and~\sevenxfortyfour\ one infers
that
$$
   w^{m-\widehat\mu}(x)^{-1}\Bigl[
   \partial_\lambda+A_\lambda^{({m-\widehat\mu})}(x)\Bigr]
   w^{m-\widehat\mu}(x)
   =w^m(x)^{-1}\Bigl[
   \partial_\lambda+A_\lambda^{(m)}(x)\Bigr]w^m(x),
\eqn\sevenxfortyseven
$$
for $x\in h(m-\widehat\mu)\cap h(m)$. Namely, the interpolated gauge
potentials defined from a side of the hypercube~$h(m-\widehat\mu)$
and defined from a side of~$h(m)$ coincide on the intersection
when~$\lambda\neq\mu$. For $\lambda=\mu$, the component may jump
across the intersection~[\GOC], but this causes no problem for our
purpose. The interpolation for the ghost field is obtained by
setting~$g(n)=\exp[\lambda c(n)]$ in the interpolation formula for
the gauge transformation~$g(x)$ in~ref.~[\GOC]. This gives the smooth
ghost field (which is also Lie algebra valued) throughout the whole
lattice.

\noindent
(iii)~The smoothness and locality. The interpolated gauge
potential~$A_\lambda(x)$ and the ghost field~$c(x)$ are smooth
functions of link variables (and of the gauge transformation function)
residing nearby the point~$x$. The smoothness (for the perturbative
configurations) and the locality are manifest from the explicit
expressions for~$A_\lambda^{(m)}(x)$ and for~$g(x)$ in~ref.~[\GOC]. In
fact, in this case, the relation is ultra-local.

\noindent
(iv)~The correct continuum limit. In the classical continuum limit,
$a\to0$, the interpolated gauge potential~$A_\mu(x)$ and the ghost
field~$c(x)$ reduce (for smooth configurations) to the gauge potential
and the ghost field in the continuum theory. From~eq.~\sevenxforty,
we have
$$
\eqalign{
   {\cal P}\exp\biggl[\int\nolimits_0^1du\,A_\mu
   (n+(1-u)\widehat\mu)\biggr]
   &=w^m(n)^{-1}u_{n,n+\widehat\mu}^mw^m(n+\widehat\mu)
\cr
   &=U(n,\mu),
\cr
}
\eqn\sevenxfortyeight
$$
where we have used the definition of the link variable in the complete
axial gauge~$u_{n,n+\widehat\mu}^m$~[\LUSCHER].\foot{In fact, from the
formulas of ref.~[\GOC], it can be seen that $A_\mu(x)$~is constant
along the link, $A_\mu(x)=A_\mu(n)$ for~$x\in[n,n+\widehat\mu]$ where
$U(n,\mu)=\exp A_\mu(n)$.} This is nothing but the conventional
expression that one assumes in the classical continuum limit,
eq.~\twoxtwentyone. For the interpolated ghost field, the formula
in~ref.~[\GOC] shows that $c(x=n)=c(n)$.

\noindent
(v)~The constant ghost field. From the formula in~ref.~[\GOC], it is
easy to see that the constant ghost field on the {\it sites\/} induces
the constant interpolated ghost field,
$$
   c(n)=c={\rm const.}\Rightarrow c(x)=c={\rm const}.
\eqn\sevenxfortynine
$$

Now we can write down the nontrivial local solution to the consistency
condition~\onexthree\ which satisfies eq.~\sevenxtwentytwo\ and the
assumptions~(I) and~(II) in terms of the {\it interpolated fields}. It
is given by
$$
\eqalign{
   {\cal A}&=-{\epsilon_H\over24\pi^2}\sum_n\int_{h(n)}d^4x\,
   \biggl\{
   \varepsilon_{\mu\nu\rho\sigma}\tr c^{(\alpha)}(x)
   \partial_\mu\biggl[A_\nu^{(\alpha)}(x)
   \partial_\rho A_\sigma^{(\alpha)}(x)
   +{1\over2}A_\nu^{(\alpha)}(x)
   A_\rho^{(\alpha)}(x)A_\sigma^{(\alpha)}(x)\biggr]
\cr
   &\qquad\qquad\qquad\quad
   +\varepsilon_{\mu\nu\rho\sigma}c^{U(1)_\beta}(x)
   \partial_\mu A_\nu^{U(1)_\beta}(x)
   \partial_\rho A_\sigma^{U(1)_\beta}(x)
\cr
   &\qquad\qquad\qquad\quad
   +\varepsilon_{\mu\nu\rho\sigma}c^{U(1)_\beta}(x)
   \tr\partial_\mu\biggl[A_\nu^{(\alpha)}(x)
   \partial_\rho A_\sigma^{(\alpha)}(x)
   +{2\over3}A_\nu^{(\alpha)}(x)
   A_\rho^{(\alpha)}(x)A_\sigma^{(\alpha)}(x)\biggr]
\cr
   &\qquad\qquad\qquad\quad
   +2\varepsilon_{\mu\nu\rho\sigma}\tr\Bigl[c^{(\alpha)}(x)
   \partial_\mu A_\nu^{(\alpha)}(x)\Bigr]
   \partial_\rho A_\sigma^{U(1)_\beta}(x)\biggr\}.
\cr
}
\eqn\sevenxfifty
$$
In this expression, $h(n)$~is the hypercube whose origin is the
site~$n$. Note that this is a functional of the link
variable~$U(n,\mu)$ and the ghost field~$c(n)$, through the
interpolation formulas of~ref.~[\GOC].

It is easy to see that eq.~\sevenxfifty\ satisfies the consistency
condition~\onexthree, because the BRS transformation of the
interpolated fields~\sevenxfortyfive\ has an identical form as that
of the continuum theory and eq.~\sevenxfifty\ has {\it formally\/} an
identical form as the gauge anomaly in the continuum
theory~\twoxtwentyeight. More precisely, we need to perform an
integration by parts within each hypercube to show~eq.~\onexthree.
Then the transverse continuity~(ii) guarantees that contributions from
a boundary of hypercubes cancel each other. This solution
eq.~\sevenxfifty\ is moreover $\delta_B$-nontrivial: From the
property~(iv) of the interpolation, in the classical continuum limit
(assuming background fields are smooth), eq.~\sevenxfifty\ reproduces
the gauge anomaly in the continuum theory~\twoxtwentyeight\ which is
BRS nontrivial. In other words, if eq.~\sevenxfifty\ is
$\delta_B$-trivial, there exists a local functional~${\cal B}$ on
lattice such that~${\cal A}=\delta_B{\cal B}$. Then the classical
continuum limit of~${\cal B}$, which is a local functional in the
continuum theory, cancels the gauge anomaly in the continuum theory.

The nontrivial solution~\sevenxfifty\ manifestly fulfills the
condition~\sevenxtwentytwo\ from the properties~(v) and~(ii) of the
interpolation. From the above arguments, it is also clear that
eq.~\sevenxfifty\ satisfies the assumption~(I) (within the
perturbative region~\twoxnine) and~(II).

The existence of the nontrivial solution ${\cal A}$~\sevenxfifty,
which satisfies eq.~\sevenxtwentytwo\ and the assumptions~(I)
and~(II), shows the existence of the unique nontrivial
sequence~$\widetilde{\cal A}_\ell$~\sevenxtwentythree. The expansion
of ${\cal A}$~\sevenxfifty\ in powers of the gauge
potential~\twoxeight\ gives the unique
sequence~$\widetilde{\cal A}_\ell$. Note that when the anomaly in the
continuum is canceled, the anomaly~${\cal A}$~\sevenxfifty\ vanishes.
Therefore the above procedure gives $\widetilde{\cal A}_\ell=0$ for
all~$\ell$. This is consistent with the conclusion in the preceding
subsection.

\chapter{Conclusion}
In this paper, we have studied the gauge anomaly~${\cal A}$ defined on
a 4-dimensional infinite lattice while keeping the lattice spacing
finite. We assumed that (I)~${\cal A}$~depends smoothly and locally on
the gauge potential, (II)~${\cal A}$~reproduces the gauge anomaly in
the continuum theory, and (III) $U(1)$~gauge anomalies have the
topological property. We have then shown that ${\cal A}$ can always be
removed by local counterterms order by order in powers of the gauge
potential: The unique exception is proportional to the anomaly in the
continuum theory. This implies that the anomaly cancellation condition
in lattice gauge theory is identical to that of the continuum theory.

As we have shown, the gauge anomaly in the formulation based on the
Ginsparg-Wilson Dirac operator satisfies the necessary prerequisites
for our result (at least for a particular choice of the integration
measure) and thus our theorems are applicable (at least in the
perturbative region in which a parameterization of the admissible
space in terms of the gauge potential is possible). Unfortunately, the
gauge anomaly ${\cal A}$ appearing in formulations based on the more
familiar Wilson Dirac operator or on the Kogut-Susskind Dirac operator
is not local, although these Dirac operators themselves are
{\it ultra-local}. For these operators, the chiral gauge symmetry is
broken at tree level, and as a result the anomaly is given
as~${\cal A}=\Tr({\rm explicit\ breaking\ term})\times%
({\rm propagator})$. The (massless) propagator in this expression
breaks the locality. (In the classical continuum limit, locality is
restored and ${\cal A}$~reproduces the gauge anomaly in the continuum
theory~[\AOK,\COS].)

Let us discuss possible extensions of the results in this paper. The
most severe limitation of our result for nonabelian theories is that
it holds only in an expansion in powers of the gauge potential. An
interesting observation related to this is that the expansion of the
anomaly density $a(n)$ (${\cal A}=\sum_n a(n)$) $a(n)=%
\sum_{\ell=1}^\infty a_\ell(n)$ in powers of the gauge potential has a
{\it finite\/} radius of convergence. This follows from the smoothness
and the locality of the anomaly~${\cal A}$ which we have assumed. If
these hold for the admissible configurations~\twoxone, the radius of
convergence of this series is given by the right hand side
of~eq.~\twoxnine. Another interesting point is that the compact
solution~\sevenxfifty\ is smooth and local at least in the
perturbative region~\twoxnine. These observations suggest that our
result is valid beyond the expansion with respect to the gauge
potential, at least within the perturbative region. What is not clear
at present is a convergence of the {\it individual\/}
series~$\widetilde{\cal A}=%
\sum_{\ell=1}^\infty\widetilde{\cal A}_\ell$ and~${\cal B}=%
\sum_{\ell=1}^\infty{\cal B}_\ell$ in~eq.~\sevenxtwentythree.

By using similar arguments as above, it seems straightforward to
classify general topological fields on a 4-dimensional infinite
lattice (which is a nonabelian analogue of the theorem~\fivexfiftysix)
at least to all orders in powers of the gauge potential. According to
the result of~ref.~[\LUSCH], this analysis is relevant for the
existence of an exactly gauge invariant formulation of anomaly-free
{\it two}-dimensional chiral gauge theories. For
{\it four}-dimensional chiral gauge theories, we have to generalize
the covariant Poincar\'e lemma~\fivexeleven\ to a 6-dimensional
lattice. This generalization would be straightforward, although the
proof may become quite cumbersome.

The restriction to the perturbative region~\twoxnine\ for nonabelian
theories is due to a complicated structure of the admissible
space~\twoxone. If it is possible to parameterize the admissible space
in terms of the gauge potential,\foot{However the existence of the
global obstruction~[\BAER] shows that this must be in general
impossible.} as in the abelian case, the restriction may be relaxed.
It is highly plausible that the ``rewinding'' technique of~ref.~[\HER]
is useful in this context.

Our results are not yet ``realistic'' because these are for infinite
lattice size. For the abelian case $G=U(1)$, it has been shown~[\LUSC]
that the anomaly cancellation works even for a finite periodic
lattice. To generalize the argument in~ref.~[\LUSC] for nonabelian
theories, we have to understand first the structure of the admissible
space on a finite lattice (see the above remark). On the other hand,
another reason why our proof is valid only for an infinite lattice is
that our proof of the algebraic Poincar\'e lemma~\fivexone\ relys on
the Poincar\'e lemma of~ref.~[\LUS], which guarantees the triviality
of the de~Rham cohomology on an infinite lattice. Physically, one
expects that the $d$-cohomology on local functions of gauge and ghost
fields is independent on the possibly nontrivial de~Rham cohomology on
a finite lattice, because the dependence is local. It is thus highly
desirable to show the algebraic Poincar\'e lemma in a way being
independent of the de~Rham cohomology. In the continuum theory, this
is actually possible~[\BRA,\DUB].

In this paper, we adopted a ``classical'' algebraic viewpoint based
on the Wess-Zumino consistency condition. In the continuum theory,
the algebraic approach to the anomaly has a close relationship to a
higher dimensional theory~[\ALV]. It seems very important to
investigate such a relationship in the context of lattice gauge
theory. In fact, there are some indications that such a
relation exists~[\LUSCH,\AOY,\ADAM].

Finally, let us remark on the physical implications of these analyses.
After all, even if a local counterterm which makes the effective
action gauge invariant exists (for anomaly-free cases), the
implementation of gauge invariance requires a fine tuning of
parameters in the counterterm which is highly unnatural. One might
thus be tempted to apply the mechanism of~ref.~[\FOR] which dynamically
restores the gauge invariance. However, for the mechanism
of~ref.~[\FOR] to work, the gauge breaking~${\cal A}$ (with the ghost
field is replaced by a logarithm of the gauge transformation field)
has to be ``small.'' In particular, if ${\cal A}\neq\delta_B{\cal B}$
for a local functional~${\cal B}$, the effective lagrangian for the
gauge transformation field is given by a lattice analogue of the
Wess-Zumino lagrangian which modifies the physical content (thus it
cannot be regarded as ``small''). Therefore, the study of the gauge
anomaly on lattice is important also in order to examine the necessary
condition for the mechanism of~ref.~[\FOR]. In this respect, it seems
interesting to study the locality (in a four dimensional sense) of the
gauge anomaly appearing in the overlap formulation with the
{\it Brillouin-Wigner phase convention}, in connection with the
result of~ref.~[\GOL].

The author has greatly benefited from correspondence with T.~Fujiwara,
Y.~Kikukawa and K.~Wu and from discussions with P.~Hern\'andez. He is
particularly grateful to M.~L\"uscher for various helpful remarks,
without which this work would not have been completed.
The author is also indebted to A. Grassi and to Y. Shamir who
independently explained to him that the statement concerning
$U(1)$~gauge anomalies, which was made in the previous version of this
paper, was not correct in general.

\Appendix{A}
Here we summarize our notation and the convention. Throughout this
paper, we consider the 4-dimensional infinite lattice~$Z^4$. The sites
of the lattice are denoted by $n$, $m$, etc. The lattice spacing is
taken to be unity~$a=1$ unless otherwise stated. The Greek
letters~$\mu$, $\nu$, etc.\ denote the Lorentz indices which run from
1 to~4. $\widehat\mu$ stands for the unit vector in direction~$\mu$.
For Lorenz indices, the summation over repeated indices is always
understood. The Levi-Civita symbol is defined
by~$\varepsilon_{\mu\nu\rho\sigma}=\varepsilon_{[\mu\nu\rho\sigma]}$
and~$\varepsilon_{1234}=1$.

The forward and the backward difference operators are respectively
defined by
$$
   \Delta_\mu f(n)=f(n+\widehat\mu)-f(n),\qquad
   \Delta_\mu^*f(n)=f(n)-f(n-\widehat\mu).
\eqn\axone
$$
The symbol~$\partial_\mu$ is reserved for the standard
{\it derivative}.

$H=R$ or~$L$ stands for the chirality of a Weyl fermion, and we set
$\epsilon_R=+1$ and~$\epsilon_L=-1$.

$G=\prod_\alpha G_\alpha$~is the (compact) gauge group where
$G_\alpha$ denotes a simple group or a $U(1)$ factor. The Greek
indices $\alpha$, $\beta$, etc.\ are used to label each factor group.
$T^a$~stands for the representation matrix of the Lie algebra,
$[T^a,T^b]=if^{abc}T^c$. The summation over repeated group
indices~$a$, $b$, etc.\ is always understood.

$U(n,\mu)$~is the link variable on the link that connects the lattice
sites $n$ and~$n+\widehat\mu$. For the abelian gauge group~$G=U(1)^N$,
we parameterize the link variable by the gauge potential
as~$U^a(n,\mu)=\exp A_\mu^a(n)$. In this case, the superscript~$a$
distinguishes each $U(1)$ factor. The abelian field strength is
defined by
$$
   F_{\mu\nu}^a(n)=\Delta_\mu A_\nu^a(n)-\Delta_\nu A_\mu^a(n).
\eqn\axtwo
$$
We {\it never\/} use this symbol~$F_{\mu\nu}^a$ to indicate the
nonabelian field strength. The plaquette variable is defined by
$$
   P(n,\mu,\nu)=U(n,\mu)U(n+\widehat\mu,\nu)
   U(n+\widehat\nu,\mu)^{-1}U(n,\nu)^{-1}.
\eqn\axthree
$$

\Appendix{B}
In this appendix, we show the calculation of the ``Wilson line''~$W'$ 
in~eq.~\twoxseventeen. From eqs.~\twoxeleven\ and~\twoxseventeen,
$W'$~is given by
$$
   W'=\exp\biggl[\epsilon_H\int\nolimits_0^1dt\,\int\nolimits_0^1ds\,
   \Tr P_t(s)\bigl[\partial_sP_t(s),\partial_tP_t(s)\bigr]\biggr],
\eqn\bxone
$$
where $P_t(s)=P_H|_{U\to U_t(s)}$ and we explicitly indicated
$s$-dependences defined through~eq.~\twoxthirteen. If we introduce the
transporting operator~$Q_t(s)$ for each~$s$ by $\partial_tQ_t(s)=%
\bigl[\partial_tP_t(s),P_t(s)\bigr]Q_t(s)$ and $Q_0(s)=1$, we have
$$
   P_t(s)=Q_t(s)P_0(s)Q_t(s)^\dagger,
\eqn\bxtwo
$$
(note that $Q_t(s)$ is unitary). Substituting this into eq.~\bxone,
and after some calculation, we have
$$
   W'=\exp\biggl\{\epsilon_H\int\nolimits_0^1dt\,\int\nolimits_0^1ds\,
   \bigl[\partial_s\Tr P_0(s)Q_t^\dagger(s)\partial_tQ_t(s)
   -\partial_t\Tr P_0(s)Q_t^\dagger(s)\partial_sQ_t(s)\bigr]\biggr\}.
\eqn\bxthree
$$
We then apply the Stokes theorem to this 2-dimensional integration.
This yields
$$
   W'=\exp\biggl[\epsilon_H\int\nolimits_0^1dt\,
   \Tr P_0(1)Q_t^\dagger(1)\partial_tQ_t(1)
   -\epsilon_H\int\nolimits_0^1ds\,
   \Tr P_0(s)Q_t^\dagger(s)\partial_sQ_t(s)\Bigr|_{t=0}^{t=1}\biggr].
\eqn\bxfour
$$
However, from eq.~\bxtwo,
$$
   \Tr P_0(1)Q_t^\dagger(1)\partial_tQ_t(1)
   =\Tr P_0(1)Q_t^\dagger(1)\bigl[\partial_tP_t(1),P_t(1)\bigr]
   Q_t(1)=0,
\eqn\bxfive
$$
and $\partial_sQ_0(s)=0$ because $Q_0(s)=1$. Therefore
$$
   W'=\exp\biggl[-\epsilon_H\int\nolimits_0^1ds\,
   \Tr P_0(s)Q_1^\dagger(s)\partial_sQ_1(s)\biggr].
\eqn\bxsix
$$
Noting that $P_0(s)\partial_sP_0(s)P_0(s)=0$ and $P_1(s)=P_0(s)$
(recall that $U_1(s)=U_0(s)$), it can be confirmed that
eq.~\bxsix\ is equal to
$$
   W'=\exp\biggl\{-\epsilon_H\int\nolimits_0^1ds\,
   \Tr\Bigl[1-P_0(s)+P_0(s)Q_1(s)\Bigr]^{-1}\partial_s
   \Bigl[1-P_0(s)+P_0(s)Q_1(s)\Bigr]\biggr\}.
\eqn\bxseven
$$
This proves eq.~\twoxseventeen.

\refout
\bye